\documentclass[twocolumn]{autart}    

\usepackage{graphicx}
\usepackage{enumerate} 
\usepackage{amsmath} 
\usepackage{amssymb} 
\usepackage{subfigure}
\usepackage{algorithm,algorithmicx,algpseudocode}
\usepackage{multirow}
\usepackage{booktabs}
\usepackage{array}
\usepackage{xcolor}

\usepackage{indentfirst}
\makeatletter
\renewcommand\normalsize{
	\@setfontsize\normalsize\@xpt\@xiipt
	\abovedisplayskip 5\p@ \@plus3\p@ \@minus3\p@
	\abovedisplayshortskip \z@ \@plus3\p@
	\belowdisplayshortskip 5\p@ \@plus3\p@ \@minus3\p@
	\belowdisplayskip \abovedisplayskip
	\let\@listi\@listI}
\makeatother
\newtheorem{theorem}{Theorem}

\newtheorem{definition}{Definition}

\newtheorem{lemma}{Lemma}
\newtheorem{proposition}{Proposition}
\newtheorem{remark}{Remark}

\graphicspath{{./}}         

\begin{document}

\begin{frontmatter}

\title{Hybrid variable monitoring: An unsupervised process monitoring framework with binary and continuous variables\thanksref{footnoteinfo}} 

\thanks[footnoteinfo]{This paper was not presented at any IFAC 
meeting. Corresponding author: Donghua Zhou, Maoyin Chen.}

\author[Tsinghua]{Min Wang}\ead{m-wang18@mails.tsinghua.edu.cn}, 
\author[ShandongTechnology,Tsinghua]{Donghua Zhou}\ead{zdh@tsinghua.edu.cn},               
\author[Tsinghua]{Maoyin Chen}\ead{mychen@tsinghua.edu.cn}

\address[Tsinghua]{Department of Automation, Tsinghua University, Beijing 100084, China}                                               
\address[ShandongTechnology]{College of Electrical Engineering and Automation, Shandong University of Science and Technology, Qingdao, 266590, China}             


\begin{abstract}                          
Traditional process monitoring methods, such as PCA, PLS, ICA, MD \textit{et al.}, are strongly dependent on continuous variables because most of them inevitably involve Euclidean or Mahalanobis distance. With industrial processes becoming more and more complex and integrated, binary variables also appear in monitoring variables besides continuous variables, which makes process monitoring more challenging. The aforementioned traditional approaches are incompetent to mine the information of binary variables, so that the useful information contained in them is usually discarded during the data preprocessing. To solve the problem, this paper focuses on the issue of hybrid variable monitoring (HVM) and proposes a novel unsupervised framework of process monitoring with hybrid variables including continuous and binary variables. HVM is addressed in the probabilistic framework, which can effectively exploit the process information implicit in both continuous and binary variables at the same time. In HVM, the statistics and the monitoring strategy suitable for hybrid variables with only healthy state data are defined and the physical explanation behind the framework is elaborated. In addition, the estimation of parameters required in HVM is derived in detail and the detectable condition of the proposed method is analyzed. Finally, the superiority of HVM is fully demonstrated first on a numerical simulation and then on an actual case of a thermal power plant.
\end{abstract}

\end{frontmatter}

\section{Introduction}
\label{Introduction}
Process monitoring is indispensable because it is the premise and guarantee for the safe and stable running of industrial systems \cite{2007Actuator,2009Reconstruction,2013GeReview,chen2018deep,2021MoniNet,SiYabin2021TIE}. In recent decades, a large number of data-driven approaches have been proposed for process monitoring \cite{1996Identification,1995Disturbance,2004Nonlinear,2010Geometric,2014Data,Yin2015Data,Shangj2017Automatica,ChenMaoyTAC,wang2020anomaly,ZHAO2021109298}. However, most of them are highly based on continuous variables because they can't avoid involving Euclidean or Mahalanobis distance and can't be utilized for hybrid variables (containing continuous and binary variables) \cite{wang2020anomaly}.

Among data-driven methods, principal component analysis (PCA) has received continuous attention once it was applied in process monitoring due to its effectiveness of data dimensionality reduction \cite{1991Multivariate,1996Identification}. Based on PCA, dynamic PCA (DPCA) adopted the technology of time lag shift to construct augmented matrix to mine time-related information \cite{1995Disturbance}.  Considering slowly changing in normal process, recursive PCA (RPCA) was proposed for adaptive process monitoring \cite{2000Recursive}. In order to capture nonlinear property, kernel PCA (KPCA) was developed \cite{B2008Nonlinear,2004Nonlinear}. Unlike PCA, partial least squares (PLS) and its variants pay much attention to quality-related fault \cite{1998RecursivePLS,2010Geometric}. To weaken the Gaussian hypothesis, independent component analysis (ICA) was proposed for process monitoring \cite{2004Statistical}. The Mahalanobis distance (MD) can also be directly used for process monitoring \cite{2019Incipient}. As understanding of the fault initiation becomes more and more thorough, the moving window methods also be proposed for incipient fault detection \cite{2016Moving,Shangj2017Automatica,QinYihao2019Data}. Considering the practical applicability in industrial processes, a large number of improved methods have been developed for multimode and nonstationary monitoring \cite{2018ZhangTII,2018Multimode,Zhao2019Fault}.

The aforementioned methods have made remarkable achievements in process monitoring, but almost all methods are based on Euclidean or Mahalanobis distance and are highly dependent on continuous variables. However, the practical industrial processes sometimes have not only continuous variables, but also binary variables which may carry some useful information for process monitoring \cite{wang2020anomaly} and are usually deleted in the data preprocessing \cite{2017Data}. For hybrid variables, Langseth \textit{et al.} used hybrid Bayesian networks for estimating human reliability \cite{2009Inference}. Aguilera \textit{et al.} developed the naïve Bayes (NB) and tree augmented naïve Bayes (TAN) models and applied to species distribution \cite{2010Hybrid}. Zhu \textit{et al.} considered the mixture of continuous and discrete variables in semantic model \cite{2012Propagation}. Talvitie \textit{et al.} introduced a related model through employing an adaptive discretization approach for structure learning in Bayesian networks when there are both continuous and discrete variables \cite{2019Learning}. Recently, Wang \textit{et al.} utilized continuous and binary (two-valued) variables to detect the abnormalities of thermal power plant for the first time \cite{wang2020anomaly}. Then a more effective anomaly monitoring model named feature weighted mixed naive Bayes model (FWMNBM) was developed \cite{MinWang2021JASfinal}. 

However, the hybrid variable approaches mentioned above are supervised methods and require both normal and fault data during training. Unfortunately, the systems in actual industrial processes are running without fault in most time, and the determinations of fault samples requires repeated research and careful discussion by experts, which are time-consuming and costly. So that the healthy state samples are usually available and it is difficult to collect sufficient fault instances, which is one of the reasons why monitoring methods only based on normal working condition data, such as PCA, PLS, ICA \textit{et al.}, have attracted much attention. Process monitoring methods with hybrid variables only based  on healthy state data are very urgently. Therefore, this paper focuses on hybrid variable process monitoring and proposes a novel unsupervised framework of process monitoring with hybrid variables named HVM which can simultaneously capture the process information of both continuous and binary variables. The main contributions are summarized as follows: 
\begin{enumerate}[(1)]
	\item {The article firstly focuses on hybrid variable monitoring only based on healthy state data. And a novel unsupervised framework of process monitoring with hybrid variables (continuous and binary variables) named HVM is proposed.}
	\item {Under the unsupervised framework, the statistics and the monitoring strategy suitable for hybrid variables are firstly defined and the physical explanation behind the framework is elaborated. In addition, the expressions of parameters are derived in detail and the detectable condition is analyzed.}
	\item {The effectiveness and efficiency of the proposed method is fully demonstrated first on a numerical simulation and then on a practical fan system of ultra-supercritical power plant.}
\end{enumerate}

The remainder of the paper is organized as follows. The problem formulation and motivation are described in detail in Section \ref{Problemreformulation}. The framework of hybrid variable monitoring is introduced in Section \ref{Hybridvariablemonitoringframework}. In Section \ref{ParametersEstimation}, parameters learning and corresponding derivation are described. The fault form of hybrid variables is defined and the detectable condition is analyzed in Section \ref{Detectabilityanalysis}. In Section \ref{Casestudies}, the effectiveness and efficiency of proposed framework is verified. Finally, conclusions are given in Section \ref{Conclusion}.

\section{Problem formulation and motivation}
\label{Problemreformulation}

With industrial processes becoming more and more complex and integrated, binary variables also appear in monitoring variables. For example, in Zhejiang Zheneng Zhongmei Zhoushan Coal and Electricity Co., Ltd. (Zhoushan Power Plant), Zhejiang Province, China, the number of monitoring variables in the No.1 power unit is about $17380$, in which the number of binary variables among them is as many as $8820$ \cite{wang2020anomaly}. In the fan system of No.1 power unit, $260$ continuous variables and $495$ binary variables are collected, where the number of binary variables is more than that of continuous variables \cite{MinWang2021JASfinal}. The appearance of binary variables makes traditional monitoring approaches no longer applicable and process monitoring with hybrid variables more intractable. The binary variables are usually discarded during the data preprocessing because the traditional approaches mostly have applied Euclidean or Mahalanobis distance which can't be used to describe binary variables \cite{2017Data}.  However, binary variables may carry some useful information for process monitoring \cite{wang2020anomaly,MinWang2021JASfinal}.

The issue of supervised classification with hybrid variables has been paid attention to and investigated in other fields \cite{2009Inference,2010Hybrid,2012Propagation,2019Learning}. In process monitoring, Wang \textit{et al.} have utilized continuous and binary variables for the anomaly detection of thermal power plant \cite{wang2020anomaly,MinWang2021JASfinal}. However, these approaches are supervised methods, which require both normal samples and fault instances to train the model. In practical processes, a lots of healthy state samples can be collected and it is difficult to obtain sufficient faulty samples. Therefore, this paper proposes a novel unsupervised framework of process monitoring with continuous and binary variables named HVM. HVM can simultaneously mine the information of both continuous and binary variables through a probabilistic framework. In HVM, the statistics of hybrid variables are computed with healthy state data and the control limit is determined by kernel density estimation (KDE) \cite{phaladiganon2013principalKDE}. Then for the arriving sample $\pmb x_a$, the statistic $s_a$ can be computed with the same way of training. The state of $\pmb x_a$ can be determined through the monitoring strategy. Finally the superiority of HVM is demonstrated through a numerical simulation and an actual case in the fan system of a thermal power plant.

\section{Hybrid variable monitoring framework}
\label{Hybridvariablemonitoringframework}
\subsection{Off-line statistics}
\label{Offlinestatistics}
Training data $\pmb X = \left\lbrace \pmb x_i\right\rbrace _{i=1}^{n} $ are sampled under normal operating condition with $n$ samples. $\pmb x_i \in \mathbb{R}^d$ is the $i$th instance and contains $d~(d=d_b+d_c)$ features where $d_b$ binary features and $d_c$ continuous features are respectively collected. Let $x^j$ be the $j$th variable.  $j_b$ and $j_c$ mean the $j_b$th and $j_c$th variable of binary variables and continuous  variables respectively. When the system is running in a steady state, the monitoring data tends to be stationary and with no trends \cite{ZHAO2021109298}. Then the following assumptions are introduced.
\begin{assum}
	\label{assumptionGaussiandistribution}
	If $x^j$ is a continuous variable (denoted as $x^{j_c}$), we suppose it obeys Gaussian distribution under normal condition, that is \cite{wang2020anomaly} 
	\begin{align}\label{continuousConditionalProbability}
		P_c( {x^{j_c}};\pmb \theta^{j_c}) = \mathcal N(x^{j_c};\mu^{j_c},\sigma^{j_c}),
	\end{align} 
	where $\pmb \theta^{j_c}=\{\mu^{j_c},\sigma^{j_c}\}$, $\mathcal N(x^{j_c};\mu^{j_c},\sigma^{j_c})$ is the probability density function (pdf) defined as $\mathcal N(x^{j_c};\mu^{j_c},\sigma^{j_c}) =$ $ {{(2\pi)^{-1/2} (\sigma^{j_c})^{-1}} }\exp ( - {{{{\left( {{x^{j_c}} - {\mu^{j_c}}} \right)}^2}}} {{2^{-1}(\sigma^{j_c})^{-2}}} )$, ${\mu^{j_c}}$ and $(\sigma^{j_c})^2$ are the mean and corresponding variance of the $j_c$th variable. 
\end{assum}

\begin{assum}
	\label{assumptionBernoullidistribution}
	If $x^j$ is a binary variable (denoted as $x^{j_b}$), the Bernoulli distribution is introduced as follows \cite{Fortuny2018Wallenius}:
	\begin{align}\label{binaryConditionalProbability}
		P_b( {x^{j_b}};\pmb \theta^{j_b}) = (\eta^j)^{x^{j_b}}{( {1 - \eta^{j_b}} )^{1 -x^{j_b}}} ,
	\end{align} 
	where $\pmb \theta^{j_b}=\{\eta^{j_b}\}$, $P_b( {x^{j_b}};\pmb \theta^{j_b})$ is the distribution series (ds), ${\eta^{j_b}}$ is the response probability which is defined as ${\eta^{j_b}}=P(x^{j_b} = 1)$.
\end{assum}

\begin{definition}
	\label{definitionfoccurrenceprobability}
In practical processes, variables are often correlated with each other. Then the occurrence probability of $\pmb x_i$ under normal condition is defined as
\begin{align}\label{theoccurrenceprobabilityundernormalcondition}
P( \pmb x_i;\pmb \theta) = \mathop \Pi \limits_{j_c = 1}^{{d_c}} P_c( {\pmb x_i^{j_c}};\pmb \theta^{j_c})^{\varphi^{j_c}}\mathop \Pi \limits_{j_b = 1}^{{d_b}} P_b( {x^{j_b}};\pmb \theta^{j_b})^{\varphi^{j_b}},
\end{align}
where $\pmb \theta=\{\pmb \theta^{j_c},\pmb \theta^{j_b},\varphi^{j_c},\varphi^{j_b}\}$, $\varphi$ means the weight of the corresponding variable.
\end{definition}

Affected by noise, there may be some outliers in data sampled under normal operating condition. Then the probability that $\pmb x_i$ belongs to $\pmb X$ can be obtained by
\begin{align}\label{Bayestheorem}
	P(\pmb x_i) = \tilde  \delta P( \pmb x_i;\pmb \theta),
\end{align} 
where $\tilde  \delta$ is the prior normal probability, which represents the confidence level of the health state data and equals to  $\tilde  \delta = 1-\delta$, $\delta$ is the significance level \cite{2018Deep}. 

\begin{proposition}
	\label{propositionpositivedecimal}	
	$\forall$ $\pmb x_i \in \pmb X$, $\exists$  a positive decimal $\alpha$ $(0 < \alpha < 1)$ to satisfy $\alpha \leq P(\pmb x_i) < 1$.
\end{proposition}

\textbf{Proof.}
	For $\pmb x_i \in \pmb X$, suppose Assumption \ref{assumptionGaussiandistribution} holds and $\varphi^{j_c}>0$ (which can be obtained by Definition \ref{definitionweight}, where $\mathcal M( x^j,x^{j'} )$ is non-negative.), then 
	\begin{align}\label{theoccurrenceprobabilityofcontinousvariables}
		0<\mathop \Pi \limits_{j_c = 1}^{{d_c}} P_c( {\pmb x_i^{j_c}};\pmb \theta^{j_c})^{\varphi^{j_c}}<1.
	\end{align}
	Since the number of training samples $n$ is an integer less than infinity, Assumption \ref{assumptionBernoullidistribution} is introduced, and $\varphi^{j_b}>0$ (which can be obtained by Definition \ref{definitionweight}.), we have
	\begin{align}\label{theoccurrenceprobabilityofbinaryvariables}
		0 \leq \mathop \Pi \limits_{j_b = 1}^{{d_b}} P_b( {x^{j_b}};\pmb \theta^{j_b})^{\varphi^{j_b}}\leq 1.
	\end{align}
	Then $0<P(\pmb x_i)<1$ for any $\pmb x_i \in \pmb X$. There must be a positive value $\varrho$ that satisfies 
	\begin{align}\label{theoccurrenceprobabilityofbinaryvariables1}
		0<\varrho \leq P(\pmb x_i)<1.
	\end{align}

	The prior normal probability $0<\tilde  \delta<1$, so that a positive decimal $0 < \alpha < 1$ can be find to satisfy $\alpha \leq P(\pmb x_i) < 1$, where $\alpha=\varrho \tilde  \delta$.  
	\qed

\begin{remark}
	$P_c( {x^{j_c}};\pmb \theta^{j_c})$ and $P_b( {x^{j_b}};\pmb \theta^{j_b})$ are probability distributions (pdf or ds), which are fitted by training data. Thus the more $\pmb x_i $ deviates from the statistical characteristics of $\pmb X$, the smaller $P( \pmb x_i;\pmb \theta)$ is and the smaller $P( \pmb x_i)$ is. 
\end{remark}	
	
\begin{definition}
	\label{definitionnaturallogarithmic}
When $P( \pmb x_i)$ of $\pmb x_i$ is obtained, then $f(\pmb x_i)$ is computed as
\begin{align}\label{naturallogarithmic}
	f(\pmb x_i) = \ln(P( \pmb x_i)),
\end{align} 
where $\ln(\cdot)$ is the natural logarithmic function.
\end{definition}

\begin{proposition}
	\label{propertysensitive}
	Compared to $P( \pmb x_i)$, $f(\pmb x_i)$ obtained in equation (\ref{naturallogarithmic}) is more sensitive to faulty instance. 	
\end{proposition}

\begin{figure}[!htp]
	\begin{center}
		\includegraphics[scale=0.6]{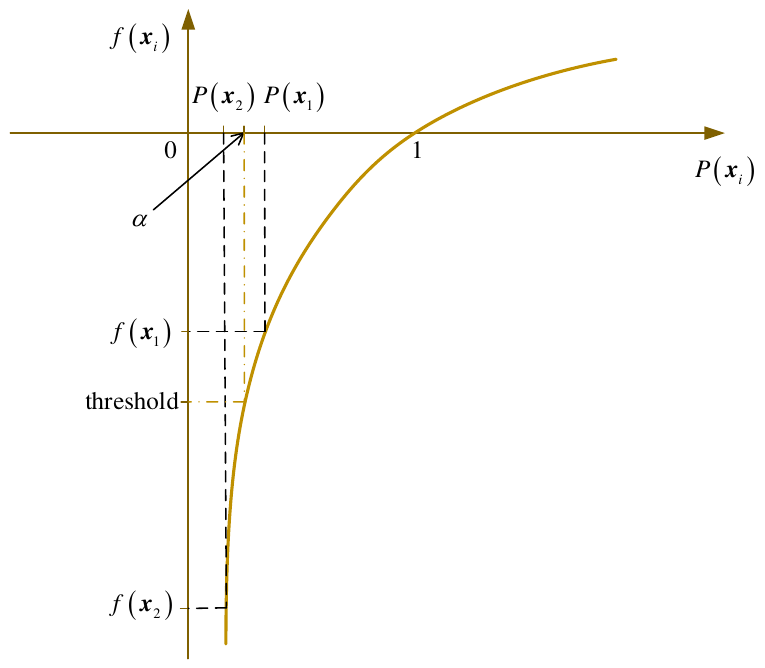}
		\caption{Transformation of the natural logarithmic function.}
		\label{Transformationnaturallogarithmicfunction}
	\end{center}
\end{figure}

\textbf{Proof.}
	According to \textbf{Proposition} \ref{propositionpositivedecimal}, a lower bound $\alpha (0<\alpha<1)$ that satisfies $P(\pmb x_i) \in [\alpha,1)$ for normal data $\pmb x_i$ can be found. For a natural logarithmic function $f(\pmb x_i)=\ln P(\pmb x_i)$, $f(\pmb x_i)$ monotonically increases and the derivative $\frac{\partial f(\pmb x_i)}{\partial P(\pmb x_i)} = \frac{1}{P(\pmb x_i)}$ always satisfy that $\frac{1}{P(\pmb x_i)} > 1$ when $0<P(\pmb x_i)<1$. Fault data $\pmb x_f$ often deviates more from the statistical characteristics of $\pmb X$ and $0<P( \pmb x_f) < \alpha$. The detection performance is mainly reflected in the recognition ability of fault in the neighborhood $U(\alpha,\epsilon)$ of $\alpha$, where $U(\alpha,\epsilon)=\{P(\pmb x_i)|\alpha - \epsilon <P(\pmb x_i)< \alpha + \epsilon\}$. Since $0<\alpha - \epsilon <P(\pmb x_i)< \alpha + \epsilon <1$, so $f(\pmb x_i)$ is more sensitive to faulty instance than $P(\pmb x_i)$. The transformation of the natural logarithmic function is shown in Fig. \ref{Transformationnaturallogarithmicfunction}. For the normal sample $\pmb x_1$ and faulty sample $\pmb x_2$, $0<P(\pmb x_2)<\alpha<P(\pmb x_1)<1$ and $\alpha-P(\pmb x_2)=P(\pmb x_1)-\alpha$. $f(\pmb x_1)$, $f(\pmb x_2)$ and \textit{threshold} are obtained from $P(\pmb x_1)$, $P(\pmb x_2)$ and $\alpha$ with natural logarithmic transformation, respectively. According to the properties of the natural logarithm function, $f(\pmb x_1)$-\textit{threshold}$<$\textit{threshold}-$f(\pmb x_2)$.
\qed

According to equation (\ref{theoccurrenceprobabilityundernormalcondition}), (\ref{Bayestheorem}) and (\ref{naturallogarithmic}), $f(\pmb x_i)$ can be written as 
\begin{align}\label{naturallogarithmicfunction}
	&f(\pmb x_i) = \ln(\tilde  \delta P( \pmb x_i;\pmb \theta)) =\ln(\tilde  \delta) + \ln(P( \pmb x_i;\pmb \theta)) \notag \\
	&=\ln{\tilde  \delta} + \ln(\mathop \Pi \limits_{j_c = 1}^{{d_c}} {P_c}( {x^{j_c}}; \pmb \theta^{j_c} )^{\varphi^{j_c}}\mathop \Pi \limits_{j_b = 1}^{{d_b}} {P_b}( {x^{j_b}}; \pmb \theta^{j_b})^{\varphi^{j_b}}).
\end{align} 

Let $\Psi=\ln(\mathop \Pi \limits_{j_c = 1}^{{d_c}} {P_c}( {x^{j_c}}; \pmb \theta^{j_c} )^{\varphi^{j_c}}\mathop \Pi \limits_{j_b = 1}^{{d_b}} {P_b}( {x^{j_b}}; \pmb \theta^{j_b})^{\varphi^{j_b}})$, it can be learned that
\begin{align}\label{naturallogarithmicfunction1}
	&\Psi= \sum_{j_b = 1}^{d_b} {\varphi^{j_b}} \ln {P_b}( {x^{j_b}}; \pmb \theta^{j_b}) +  \sum_{j_c = 1}^{d_c} {\varphi^{j_c}}\ln {P_c}( {x^{j_c}}; \pmb \theta^{j_c}).
\end{align} 

Considering equation (\ref{binaryConditionalProbability}), we have
\begin{align}\label{sumbinaryConditionalProbability}
	&\sum_{j_b = 1}^{d_b} {\varphi^{j_b}} \ln {P_b}( {x^{j_b}}; \pmb \theta^{j_b}) = \sum_{j_b = 1}^{d_b} {\varphi^{j_b}} \ln [(\eta^{j_b})^{x^{j_b}}{( {1 - \eta^{j_b}} )^{1 -x^{j_b}}}] \notag \\
	&~~~~~~~~~~~~~~=\sum_{j_b = 1}^{d_b} {\varphi^{j_b}} [{x^{j_b}} \ln (\eta^{j_b}) + (1 -x^{j_b}) \ln ( \tilde \eta^{j_b} )]
	\notag \\
	&~~~~~~~~~~~~~~=\sum_{j_b = 1}^{d_b} [{\varphi^{j_b}}{x^{j_b}}\ln \frac{\eta^{j_b}}{\tilde \eta^{j_b}}] +\sum_{j_b = 1}^{d_b} {\varphi^{j_b}} \ln \tilde \eta^{j_b},
\end{align} 
where $ \tilde \eta^{j_b}=1 - \eta^{j_b}$. According to equation (\ref{continuousConditionalProbability}), the following equation can be obtained that
\begin{align}\label{sumcontinuousConditionalProbability}
	&\sum_{j_c = 1}^{d_c} {\varphi^{j_c}} \ln {P_c}( {x^{j_c}}; \pmb \theta^{j_c}) = \sum_{j_c = 1}^{d_c} {\varphi^{j_c}} \ln \mathcal N(x^{j_c};\mu^{j_c},\sigma^{j_c}) \notag \\
	&~~~~~~~~~~~~~~~~~~~~= \sum_{j_c = 1}^{d_c} {\varphi^{j_c}} \ln [{{(2\pi)^{-1/2} (\sigma^j)^{-1}} }]\notag \\
	&~~~~~~~~~~~~~~~~~~~~+\sum_{j_c = 1}^{d_c} {\varphi^{j_c}}[ - {{{{\left( {{x^j} - {\mu^j}} \right)}^2}}}{{2^{-1}(\sigma^j)^{-2}}} ].
\end{align} 

Substituting equation (\ref{sumbinaryConditionalProbability}) and (\ref{sumcontinuousConditionalProbability}) into equation (\ref{naturallogarithmicfunction}), $f(\pmb x_i)$ is learned as
\begin{align}\label{functionxi}
	&f(\pmb x_i) =  { \pmb \tau} _i \cdot \tilde {\pmb x}_i^T + \xi _i + \varepsilon_i,
\end{align}
where ${ \pmb \tau}_i = [\vartheta^1, \ldots ,\vartheta^{j_b}, \ldots ,\vartheta^{d_b}]$, $\vartheta^{j_b}=\varphi^{j_b} \ln \frac{\eta^{j_b}}{\tilde \eta^{j_b}}$, $\varepsilon_i = \sum\limits_{j_c = 1}^{{d_c}} \varphi^{j_c} {\left[{ \ln ( {(2\pi)^{- \frac{1}{2}} (\sigma^{j_c})^{-1}} ) - \frac{1}{2} {( {\pmb x_i^{j_c}  - {\mu^{j_c} }} )^2}{(\sigma^{j_c})^{-2}}} \right] }$, $\tilde {\pmb x}_i =  [ \pmb x_i^1 , \ldots ,\pmb x_i^{j_b},$ $ \ldots , \pmb x_i^{{d_b}} ]$, $\xi _i =  \ln (1-\delta) + \sum\limits_{j_b = 1}^{{d_b}} \varphi^{j_b} {\ln \tilde \eta^{j_b}}$.

$f(\pmb x_i)$ obtained in equation \eqref{functionxi} is negative. The statistics in process monitoring are often positive, and the judgment logic is generally that the statistics of the faulty data exceed the control limit. Thus for the collected training samples $\pmb X$, the monitoring statistics $\pmb s$ are computed as 
\begin{align}\label{statisticsofX}
	\pmb s &= [s_1,\cdots,s_i,\cdots, s_n]\notag \\
	&=[f^2(\pmb x_1), \cdots , f^2(\pmb x_i), \cdots , f^2(\pmb x_n)],
\end{align}
where $s_i$ is the statistic of $\pmb x_i$.

\subsection{On-line monitoring strategy}
\label{Onlinemonitoringstrategy}
When the statistics $\pmb s$ of $\pmb X$ are obtained, the control limit $s^{\lim}$ can be got with the significance level $\delta$ by KDE \cite{phaladiganon2013principalKDE}, $\delta=0.01$ in this paper. In online detection, the statistic $s_a$ of arriving sample $\pmb x_a$ is computed by equation (\ref{functionxi}) and (\ref{statisticsofX}). Then the state of $\pmb x_a$ is determined through the monitoring strategy:
\begin{align}\label{monitoringstrategy}
	\begin{cases}
		\pmb x_a \text{ is normal,} &  \text{if } s_a < s^{\lim},\\
		\pmb x_a \text{ is faulty,} &  \text{otherwise}.
	\end{cases}
\end{align}

\begin{remark}
	Only continuous and binary variables are considered in this work. The Bernoulli distribution is introduced for binary variable which has only two values, where 0 and 1 can also denote two state such as high or low. It should be noted that the idea and the skills for binary variables in this work can be referenced to discrete variables with more than two values. Then the Bernoulli distribution should be replaced by the multinomial distribution and the subsequent processing of the model may also need to be adjusted.
\end{remark}

\section{Parameters learning}
\label{ParametersEstimation}
The model described in \ref{Offlinestatistics} mainly involves the estimation of parameters $\mu^j$, $\sigma^j$, $\eta^j$, and $\varphi^j$. 
$\mu^j$, $\sigma^j$ and $\eta^j$ can be obtained through maximum likelihood estimation (MLE) \cite{Collins2004Parameter}. 
\begin{align}
	\label{mean}
	&u^j= \sum\limits_{i = 1}^{n} x_i^j / n,\\
	\label{standarddeviation}
	&\sigma^j=\{{\sum\limits_{i = 1}^{n} {{ \left( {x_i^j - u^j} \right) }^2} }\}^{1/2} {( {n  - 1} )}^{-1/2}, 	\\
	\label{responsefunction}
	&\eta^j=\sum\limits_{i = 1}^{n} x_i^j / n ,~~~ (x_i^j \in \{0,1\}).
\end{align}

In practical process, variables are usually correlated with others and variables that are  more related to the other variables are more sensitive when abnormalities occur \cite{wang2020anomaly}. So each variable is assigned with the different feature weight $\varphi^j$ \cite{Jiang8359364}. The mutual information (MI) can capture the dependence of variables, both linear and non-linear \cite{Darbellay761290}, and is used to construct the feature weight $\varphi^j$ which is defined as follows.
	
\begin{definition}
	\label{definitionweight}
	For the $j$th variable, the weight $\varphi^j$ is defined as
	\begin{align}\label{weight}
		\varphi^j=1+{\frac{1}{d-1}} \sum\limits_{j = 1,j \ne j'}^d {\mathcal M( x^j,x^{j'} )}.
	\end{align}
	where $\mathcal M(x^j,x^{j'})$ is the MI of $x^j$ and $x^{j'}$.	
\end{definition}

If $x^j$ and $x^{j'}$ are continuous variables or both are binary variables, $\mathcal M(x^j,x^{j'})$ can be obtained by the definition of MI for continuous variables or discrete variables. However, $x^j$ and $x^{j'}$ may include both continuous and binary variables. Then the auxiliary binary variable is constructed by Definition \ref{definition2} for the continuous variable when the feature weight is computed.

\begin{definition}
	\label{definition2}
	If $x^j$ is a continuous variable, $x'^j$ is constructed as
	\begin{align}\label{clippingprocessing}
		{x'}_i^j= \textbf{[}  x_i^j>\mu^j \textbf{]},
	\end{align}
	where $\textbf{[} \cdot \textbf{]}$ is Iverson brackets. If the condition $x_i^j>\mu^j$ is true, it returns $1$, otherwise it returns $0$.
	
\end{definition}

Definition \ref{definition2} makes it possible to characterize the correlation between hybrid variables. Then  $x'^j$ instead of $x^j$ is used to compute MI. However, if $x^j$ and $x^{j'}$ are continuous variables, $\mathcal M( x^j,x^{j'} )$ and $\mathcal M( {x'}^j, {x'}^{j'} )$ are not completely equivalent. The relationship between $\mathcal M( x^j,x^{j'} )$ and $\mathcal M( {x'}^j, {x'}^{j'} )$ is shown in Theorem \ref{theoremrelationship}.

\begin{theorem}
		\label{theoremrelationship}
		If $x^j$ and $x^{j'}$ are continuous variables, $x^j$ and $x^{j'}$ obey Gaussian distributions $ \mathcal N(\mu^j,(\sigma^j)^2)$ and $\mathcal N(\mu^{j'},(\sigma^{j'})^2)$ respectively, ${x'}^j$ and ${x'}^{j'}$ are constructed by equation (\ref{clippingprocessing}), the relationship between $\mathcal M({x'}^j, {x'}^{j'})$ and $\mathcal M({x}^j, {x}^{j'})$ is
		\begin{align}\label{Theoremmutualinformation}
			&\mathcal M({x'}^j, {x'}^{j'}) =(\frac{1}{{\pi }}\arcsin \rho+ 0.5) \log (\frac{2}{{\pi }}\arcsin \rho+ 1)\notag \\
			&+ (0.5-\frac{1}{{\pi }}\arcsin \rho)\log (1-\frac{2}{{\pi }}\arcsin \rho),
		\end{align}
		where $\rho$ is the correlation coefficient of continuous variables $x^j$ and $x^{j'}$, and is expressed as $\rho =  [1-e^{-2\mathcal M(x^j, x^{j'})}]^{1/2}$.
\end{theorem}

\textbf{Proof.} See Lemma \ref{lemma1} and appendix \ref{appendix1}.\qed

\begin{lemma}
		\label{lemma1}
		\cite{TsonisProbing} For continuous variables $x^j$ and $x^{j'}$ that follow Gaussian distributions, $\mathcal M(x^j, x^{j'})$ is the MI of $x^j$ and $ x^{j'}$. $\rho$ is the correlation coefficient between $x^j$ and $ x^{j'}$. Then
		\begin{align}\label{Theoremmutualinformation1}
			\rho =  [1-e^{-2\mathcal M(x^j, x^{j'})}]^{1/2}.
		\end{align}
\end{lemma}

The Lemma \ref{lemma1} is proved in \cite{1998Predictability}.

With Definition \ref{definition2}, the MI computation of hybrid variables is transformed to that of binary variables (or constructed binary variables). $\mathcal M(x^j, x^{j'})$ is defined as
\begin{align}\label{definecomputingofMI}
	\mathcal M(x^j, x^{j'}) = \sum\limits_{{x^j},{x^{j'}}} {P( {{x^j},{x^{j'}}})\log \frac{{P( {{x^j},{x^{j'}}})}}{{P( {{x^j}})P( {{x^{j'}}})}}} ,
\end{align}
where ${x^j}, {x^{j'}}$ are binary variables or constructed binary variables. $P({x^j})$ is the probability of $x^j=\psi_{x^j}$, $\psi_{x^j}$ is the indicative coefficient ($\psi_{x^j}=1$ when $P({x^j}=1)$ is computed, and $\psi_{x^j}=0$ otherwise), $P( {{x^j},{x^{j'}}})$ is the joint probability of $x^j=\psi_{x^j}$ and $x^{j'}=\psi_{x^{j'}}$ . $P({x^j})$ ($P(x^{j'})$ can be obtained in the same way) can be computed by 
\begin{align}\label{pxjk}
	&P({x^j})=\psi_{x^j} \frac{\sum\limits_{i = 1}^{n} {x_i^j} } {n} +(1-\psi_{x^j} )(1-\frac{\sum\limits_{i = 1}^{n} {x_i^j} } {n}),
\end{align}
where $x_i^j$ is the value at time $i$ of $x^j$. 

\begin{proposition}
	\label{propositionPxjxj}
	For binary variables $x^j$ and $x^{j'}$,  $P(x^j,x^{j'})$ can be denoted as 
	\begin{align}\label{Pxjxj1denoted}
		&P(x^j,x^{j'})=P(x^{j'}=\psi_{x^{j'}})\varsigma^{\psi_{x^j}\psi_{x^{j'}}} (1-\varsigma)^{\psi_{x^{j'}}-\psi_{x^j}\psi_{x^{j'}}} \notag \\
		&~~~~~~~~\times \varsigma'^{\psi_{x^j}-\psi_{x^j}\psi_{x^{j'}}}(1-\varsigma')^{1+\psi_{x^j}\psi_{x^{j'}}-\psi_{x^j}-\psi_{x^{j'}}},
	\end{align}
	where $P( x^j = 1| x^{j'}=1) = \varsigma, P( x^j = 1| x^{j'}=0) = \varsigma' $.
\end{proposition}

\textbf{Proof.} See appendix \ref{appendixProofproposition3}.\qed

\begin{theorem}
	\label{theorempxjxj1}
	For binary variables $x^j$ and $x^{j'}$, $P( {{x^j},{x^{j'}}})$ is obtained as
	\begin{align}\label{pxjxj1}
		P(x^j,x^{j'})&=P(x^{j'}=\psi_{x^{j'}})\{ 1 - \psi_{x^j} + (2\psi_{x^j} - 1 ) \notag \\
		&\times [ \psi_{x^{j'}}{\varsigma} + ( {1 - \psi_{x^{j'}} } ){\varsigma '} ]\},
	\end{align}
	where $\varsigma = \sum\limits_{i = 1}^{n} ( {x_i^j x_i^{j'}} )(\sum\limits_{i = 1}^{n} x_i^{j'} )^{-1}$, $\varsigma' = (\sum\limits_{i = 1}^{n} {x_i^j}  - \sum\limits_{i = 1}^{n} x_i^j x_i^{j'} )$ $({n}-\sum\limits_{i = 1}^{n} x_i^{j'})^{-1}$.
\end{theorem}
\textbf{Proof.} See appendix \ref{appendixProofTheoremtheorempxjxj1}.\qed

After $\mathcal M(x^j, x^{j'})$ is estimated, the weight $\varphi^j$ of $j$th variable could be obtained through \eqref{weight}.

\begin{remark}
	When the correlation between variables is not considered, that is $\varphi^j=1$, all variables have the same weight.
\end{remark}

\section{Detectability analysis}
\label{Detectabilityanalysis}

\subsection{Fault description}
\label{Faultdescription}

In multivariate statistical process monitoring, the fault model is usually described as 
\begin{align}\label{Faulmodel}
	X^f= X + \varXi F,
\end{align}
where $\varXi$ is the fault direction vector, $F$ represents the fault
magnitude vector\cite{2009Reconstruction,Shangj2017Automatica}. The emergence of binary variables makes that the fault model described in equation (\ref{Faulmodel}) is no longer suitable. Thus the fault model of hybrid variables is defined as follows.
\begin{definition}
	\label{definition4}
	The fault model of hybrid variables (containing continuous and binary variables) is defined as 
	\begin{align}\label{Faultdescriptionequation}
		\pmb X^f= \pmb X + \pmb \varXi \circ \pmb F,
	\end{align}
	where $\pmb X$ is the healthy state data, $\pmb \varXi$ means the fault direction matrix, $\pmb F$ represents the fault magnitude matrix, $\circ$ is Hadamard product \cite{1985Matrix} which means the corresponding elements of $\pmb \varXi$ and $\pmb F$ are multiplied, $\pmb X^f$ is the fault data.
\end{definition}

\begin{remark}
	When only continuous variables are monitored, fault can be described as multiplying with the direction vector and the amplitude vector. In fact, equation (\ref{Faulmodel}) is a special case of equation (\ref{Faultdescriptionequation}).
\end{remark}

The fault model at time $i$ in Definition \ref{definition4} is 
\begin{align}\label{Faultdescriptionequation1}
	(\pmb X^f)_i= (\pmb X)_i + \pmb \varXi_i \circ \pmb F_i,
\end{align}
where $(\pmb X^f)_i, (\pmb X)_i, \pmb \varXi_i, \pmb F_i$ are $1\times d$ vectors. If there is a fault at time $i$ and it occurs on the $j$th ($1 \le j \le d$) variable, then $\pmb \varXi_i^j=1$, $\pmb F_i^j$ means the corresponding fault amplitude, otherwise $\pmb \varXi_i^j=0$ or $\pmb F_i^j=0$.  Faults can also appear on multiple variables.

\begin{remark}
	If the $j$th variable is a binary variable, the fault amplitude $\pmb F^j$ must be $1$ or $-1$. $\pmb F^j$ must be $-1$ when $(\pmb X)^j=1$, and $\pmb F^j$ must be $1$ when $(\pmb X)^j=0$.
\end{remark}

\subsection{Detectability conditions}
\label{Detectabilityconditions}
According to the monitoring strategy (\ref{monitoringstrategy}), the state of $\pmb x_a$ is judged as fault if the statistic $s_a$ exceed the control limit $s^{lim}$. The detectable condition is shown as the following Theorem.
\begin{theorem}
	\label{theorem5}
	For the arriving sample $\pmb x_a$, it can be judged to be faulty if and only if
	\begin{align}\label{pxjxj11}
		P( \pmb x_a;\pmb \theta)<\tilde \delta e^{\tilde s},
	\end{align}
	where $\tilde s=-\sqrt{s^{lim}}$, $	P( \pmb x_a;\pmb \theta)=\mathop \Pi \limits_{j_c = 1}^{{d_c}} \mathcal N(\pmb x_a^{j_c};\mu^{j_c},\sigma^{j_c})^{\varphi^{j_c}}$ $\mathop \Pi \limits_{j_b = 1}^{{d_b}} [(\eta^{j_b})^{\pmb x_a^{j_b}}{( {1 - \eta^{j_b}} )^{1 -\pmb x_a^{j_b}}} ]^{\varphi^{j_b}}$,  $\mathcal N(\pmb x_a^{j_c};\mu^{j_c},\sigma^{j_c}) = {{(2\pi)^{-1/2} (\sigma^{j_c})^{-1}} }\exp ( - {{{{\left( {{{\pmb x_a}^{j_c}} - {\mu^{j_c}}} \right)}^2}}}$ ${{2^{-1}(\sigma^{j_c})^{-2}}} )$.
\end{theorem}

\textbf{Proof.} The fault occurs if the statistic $s_a$ of $\pmb x_a$ exceeds the control limit $s^{lim}$. According to Proposition \ref{propositionpositivedecimal} and Proposition \ref{propertysensitive}, the state of $\pmb x_a$ is judged as fault when $0<P( \pmb x_a) < \alpha$. Let $\ln^2 \alpha = s^{lim}$, we have
\begin{align}\label{controllimit}
	\alpha= e^{-\sqrt{s^{lim}}}
\end{align}

Then the following inequality can be obtained
\begin{align}\label{inequality}
	0<P(\pmb x_a) < e^{\tilde s}
\end{align}
where $\tilde s=-\sqrt{s^{lim}}$. When the significance level $\delta$ is given, $\tilde \delta=1-\delta$. It can be learned that $P( \pmb x_a)<\tilde \delta e^{\tilde s}$. According to Assumption \ref{assumptionGaussiandistribution} and \ref{assumptionBernoullidistribution}, Theorem \ref{theorem5} is proved. \qed

The procedure is summarized in Algorithm \ref{EMFWMNBM}.

\begin{algorithm}
	\caption{HVM}
	\label{EMFWMNBM}
	\begin{algorithmic}
		\item[\textbf{Off-line modeling:}]
		\item[$Step$ 1] {Identify continuous and binary variables.}
		\item[$Step$ 2] {Estimate the means $\mu^j$ and the standard deviation $\sigma^j$ for each continuous variable via (\ref{mean}) and (\ref{standarddeviation}).}
		\item[$Step$ 3] {Estimate the response functions ${\eta^j}$ for each binary variable via (\ref{responsefunction}).}
		\item[$Step$ 4] {Give the significance level $\delta$ and the confidence level $\tilde  \delta = 1-\delta$. } 
		\item[$Step$ 5] {Construct ${x'}^j$ for each continuous variable $x^j$ according to Definition \ref{definition2}.}
		\item[$Step$ 6] {Estimate probability $P({x^j})$ and joint probability $P(x^j,x^{j'})$ via (\ref{pxjk}) and (\ref{pxjxj1}).} 
		\item[$Step$ 7] {Estimate MI $\mathcal M(x^j, x^{j'})$ between $x^j$ and $x^{j'}$ via (\ref{definecomputingofMI}).}
		\item[$Step$ 8] {Estimate weight $\varphi^j$ via (\ref{weight}).}
		\item[$Step$ 9] {Calculate $f(\pmb x_i)$ of $\pmb x_i$ via (\ref{functionxi}).}
		\item[$Step$ 10] {Calculate statistics $\pmb s$ of $\pmb X$ via (\ref{statisticsofX}).}
		\item[$Step$ 11] {Calculate control limit $s^{lim}$ through KDE.}
		
		\item[\textbf{On-line monitoring:}]
		\item[$Step$ 12] {Construct $\tilde {\pmb x}_a$ through the arriving sample $\pmb x_a$.}
		\item[$Step$ 13] {Calculate $\varepsilon_a$ and  $\xi _a$.}
		\item[$Step$ 14] {Calculate $f(\pmb x_a)$ of $\pmb x_a$ via (\ref{functionxi}).}
		\item[$Step$ 15] {Calculate statistic $s_a$ of $\pmb x_a$ via (\ref{statisticsofX}).}
		\item[$Step$ 16] {Determine the state of sample $\pmb x_a$ via (\ref{monitoringstrategy}).}
	\end{algorithmic}
\end{algorithm}

\section{Experimental verification}
\label{Casestudies}
In this section, the superiority of HVM is demonstrated through two cases. 

\subsection{Numerical case} 
\begin{table}[bp]	
	\scriptsize
	\centering
	\caption{The distributions of continuous variables.}
	\label{TheparametersofCVs}
	\scalebox{1.00}[1.00]{
		\begin{tabular}{m{0.2cm}<{\centering}m{1.5cm}<{\centering}m{1.5cm}<{\centering}m{1.5cm}<{\centering}m{1.5cm}<{\centering}}
			\toprule
			\multirow{3}{*}{ } & \multicolumn{2}{c}{Experiment \textrm{I}} & \multicolumn{2}{c}{Experiment \textrm{II}} \\
			
			\cmidrule(r){2-3} \cmidrule(r){4-5}
			
			&\multicolumn{1}{c}{normal($\pmb X$)} &\multicolumn{1}{c}{fault ($\pmb F$)}  &\multicolumn{1}{c}{normal ($\pmb X$)} &\multicolumn{1}{c}{fault ($\pmb X^f$)} \\
			
			\midrule
			
			$x^1$ &$\mathcal N(1.35,0.66^2)$ &$\mathcal N(0.15,0.66^2)$ &$\mathcal N(1.50,0.76^2)$ &$\mathcal N(0.55,0.55^2)$ \\
			$x^2$ &$\mathcal N(2.65,0.80^2)$ &$\mathcal N(0.05,0.78^2)$ &$\mathcal N(3.00,0.68^2)$ &$\mathcal N(2.55,1.01^2)$ \\
			$x^3$ &$\mathcal N(0.86,0.66^2)$ &$\mathcal N(0.10,0.60^2)$ &$\mathcal N(1.70,0.85^2)$ &$\mathcal N(2.20,1.00^2)$ \\
			$x^4$ &$\mathcal N(1.80,0.90^2)$ &$\mathcal N(0.15,0.89^2)$ &$\mathcal N(0.80,1.01^2)$ &$\mathcal N(1.45,0.91^2)$ \\
			$x^5$ &$\mathcal N(0.99,0.55^2)$ &$\mathcal N(0.30,0.58^2)$ &$\mathcal N(0.89,0.64^2)$ &$\mathcal N(1.30,0.55^2)$ \\
			\bottomrule	
	\end{tabular}	}
\end{table}
The fault model is considered as follows:
\begin{align}\label{Faultdescriptionnumericalcase}
	\pmb X^f= \pmb X + \pmb \varXi \circ \pmb F
\end{align}
where $\pmb X, \pmb \varXi, \pmb F, \pmb X^f \in \mathbb{R}^{n\times d}$. $\pmb \varXi=\pmb0$ in the normal working condition. $\pmb X$ contains $5$ continuous variables $(x^1, \ldots , x^5)$ and $5$ binary variables $(x^6, \ldots , x^{10})$. 4000 normal samples are generated for training. Then 4000 instances are collected for verifying the effectiveness and efficiency of the proposed model. The fault is introduced from time $2001$.  
\begin{table}[bp]
	\scriptsize
	\centering
	\caption{The parameters of binary variables.}
	\label{TheparametersofBVs}
	
	\begin{tabular}{m{0.5cm}<{\centering}m{0.4cm}<{\centering}m{0.4cm}<{\centering}m{0.4cm}<{\centering}m{0.4cm}<{\centering}m{0.4cm}<{\centering}m{0.4cm}<{\centering}m{0.4cm}<{\centering}m{0.4cm}<{\centering}}
		\toprule
		
		\multirow{4}{*}{} &\multicolumn{4}{c}{Experiment \textrm{I}} &\multicolumn{4}{c}{Experiment \textrm{II}} \\
		
		\cmidrule(r){2-5} \cmidrule(r){6-9}
		
		&\multicolumn{2}{c}{normal ($\pmb X$)} &\multicolumn{2}{c}{fault ($\pmb F$)} &\multicolumn{2}{c}{normal ($\pmb X$)} &\multicolumn{2}{c}{fault  ($\pmb X^f$)} \\
		
		\cmidrule(r){2-3} \cmidrule(r){4-5} \cmidrule(r){6-7} \cmidrule(r){8-9} 
		
		&value &ratio &value &ratio &value &ratio &value &ratio \\
		
		\midrule
		
		$x^6$ &0 &5 &0 &50 &0 &10 &1 &5 \\
		$x^7$ &0 &6 &0 &45 &0 &5 &1 &10 \\
		$x^8$ &1 &12 &0 &38 &1 &15 &0 &10 \\
		$x^9$ &1 &2 &0 &35 &1 &8 &0 &15 \\
		$x^{10}$ &1 &8 &0 &48 &1 &10 &0 &8 \\
		
		\bottomrule	
	\end{tabular}	
\end{table}
\begin{figure*}[htbp]
	\centering
	
	\subfigure[PCA ({$T^2$})]
	{
		\centering
		\includegraphics[width=2.3in]{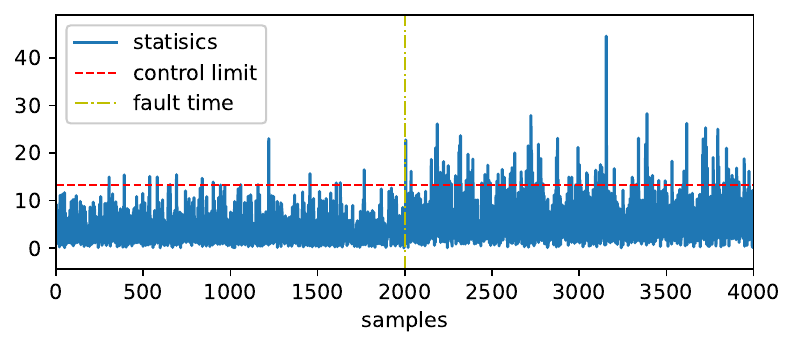}
	}%
	\subfigure[PCA ({ $Q$})]
	{
		\centering
		\includegraphics[width=2.3 in]{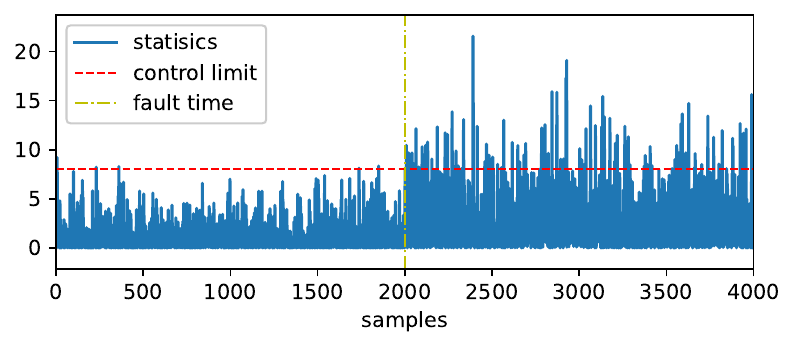}
	}%
	\subfigure[DPCA ({$T^2$})]
	{
		\centering
		\includegraphics[width=2.3in]{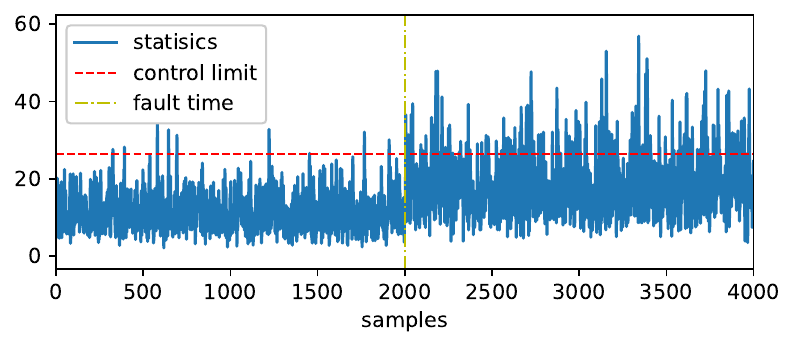}
	}%

	\subfigure[DPCA ({$Q$})]
	{
		\centering
		\includegraphics[width=2.3 in]{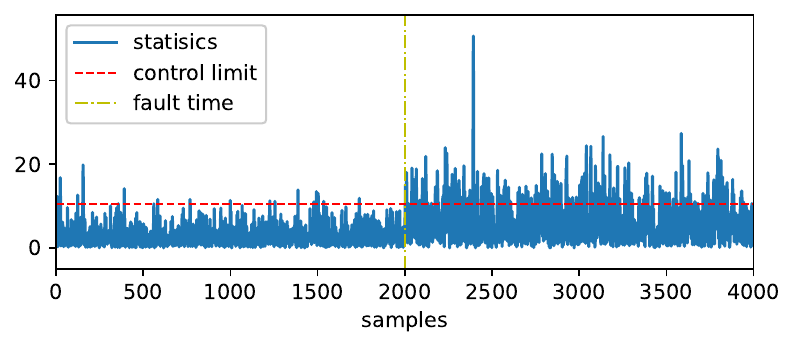}
	}%
	\subfigure[ICA ({$I^2$})]
	{
		\centering
		\includegraphics[width=2.3in]{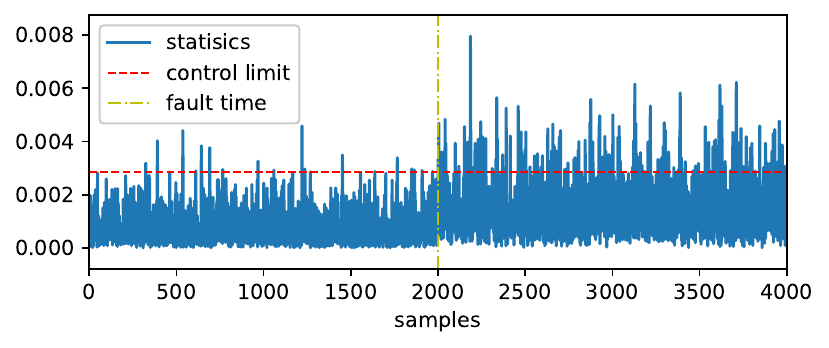}
	}%
	\subfigure[ICA ({$I_e^2$})]
	{
		\centering
		\includegraphics[width=2.3 in]{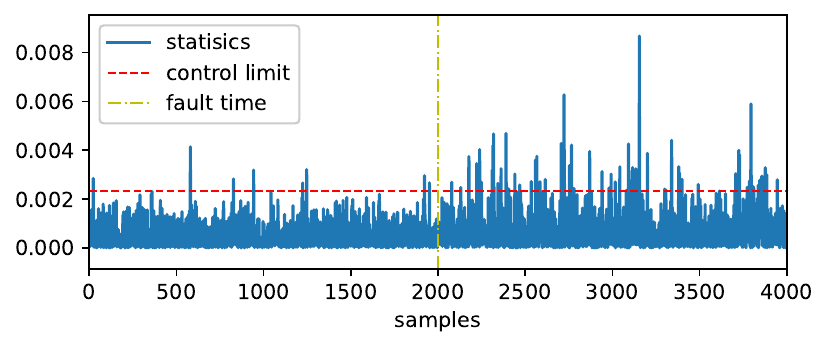}
	}%

	\subfigure[ICA ({$Q$})]
	{
		\centering
		\includegraphics[width=2.3in]{statistics_ICA2}
	}%
	\subfigure[MD]
	{
		\centering
		\includegraphics[width=2.3 in]{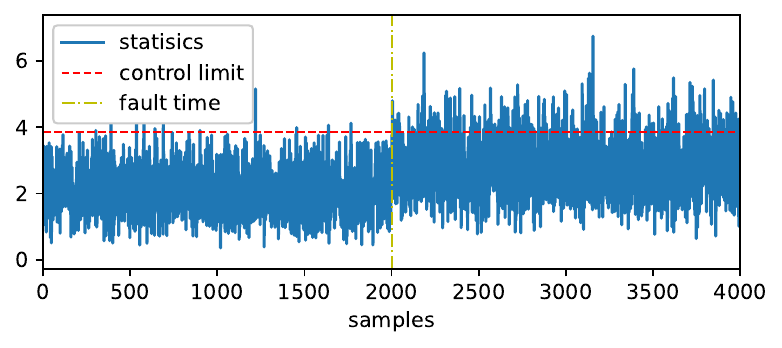}
	}%
	\subfigure[HVM]
	{
		\centering
		\includegraphics[width=2.3 in]{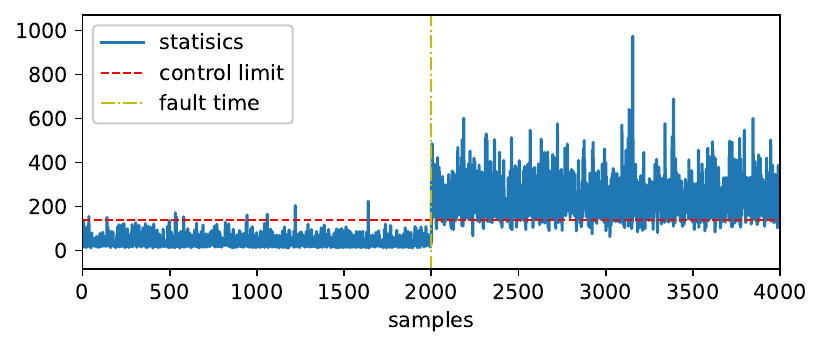}
	}%

	\centering
	\caption{Detection performance in experiment \textrm{II}.}
	\label{fault1Simu}
\end{figure*}

Two experiments are conducted.  Under normal condition in experiment \textrm{I}, the distributions of continuous variables and the values and ratios of binary variables are shown as normal ($\pmb X$) in Table \ref{TheparametersofCVs} and Table \ref{TheparametersofBVs} respectively. A fault occurred from the time 2001, the continuous variables were disturbed by the Gaussian noises whose distributions is shown as  fault ($\pmb F$)  in Table \ref{TheparametersofCVs}, the ratios of binary variables after fault arriving are listed as fault ($\pmb F$) in Table \ref{TheparametersofBVs}. In experiment \textrm{II}, the process information carried by the binary variable is increased, and the difference in the distributions of continuous variables under normal and fault conditions is narrowed. The continuous variable distributions before fault occurring are assumed as normal ($\pmb X$) of experiment \textrm{II} in Table \ref{TheparametersofCVs}, the values and ratios of binary variables under normal condition are listed as normal ($\pmb X$) of experiment \textrm{II} in Table \ref{TheparametersofBVs}. The values of binary variables after fault significantly changed which is depicted as fault ($\pmb X^f$) of experiment \textrm{II} in Table \ref{TheparametersofBVs}. In order to make it more general, random jumps are added on binary variables and the adjustment ratio is shown in Table \ref{TheparametersofBVs}. Random jump means that the value changes at a time and recovers at the next moment. The distributions of continuous variables after fault are listed as fault ($\pmb X^f$) of experiment \textrm{II} in Table \ref{TheparametersofCVs}.
\begin{table}[tp]
	\scriptsize
	\centering
	\caption{The means of FARs and FDRs in the numerical study.}
	\label{ThemeansofFARsandFDRs}
	\begin{tabular}{m{0.5cm}<{\centering}m{0.3cm}<{\centering}m{0.4cm}<{\centering}m{0.4cm}<{\centering}m{0.4cm}<{\centering}m{0.3cm}<{\centering}m{0.4cm}<{\centering}m{0.3cm}<{\centering}m{0.5cm}<{\centering}m{0.8cm}<{\centering}}
		\toprule
		\multirow{3}{*}{}
		& \multicolumn{2}{c}{PCA} & \multicolumn{2}{c}{DPCA} & \multicolumn{3}{c}{ICA} & \multicolumn{1}{c}{\multirow{3}{*}{MD}} & \multicolumn{1}{c}{\multirow{3}{*}{HVM}} \\
		
		\cmidrule(r){2-3} \cmidrule(r){4-5} \cmidrule(r){6-8}
		
		&$T^2$ &$Q$ &$T^2$ &$Q$ &$I^2$ &$I_e^2$ &$Q$ &  &  \\
		
		\midrule
		
		FAR$_{\textrm{I}}$  &0.94 &0.98 &0.92 &0.89 &0.87 &0.95 &0.90 &0.93 &0.62 \\
		FDR$_{\textrm{I}}$  &7.45 &13.77 &10.26 &35.21 &8.97 &14.08 &8.59 &16.63 &52.34 \\
		FAR$_{\textrm{II}}$  &0.83 &0.27 &0.87 &0.19 &0.72 &0.90 &0.62 &0.64 &0.60 \\
		FDR$_{\textrm{II}}$  &1.47 &2.40 &0.86 &6.47 &3.22 &4.96 &0.47 &6.01 &94.73 \\
		
		\bottomrule	
	\end{tabular}
\end{table}
\begin{figure*}[!htp]
	\begin{center}
		\includegraphics[scale=0.315]{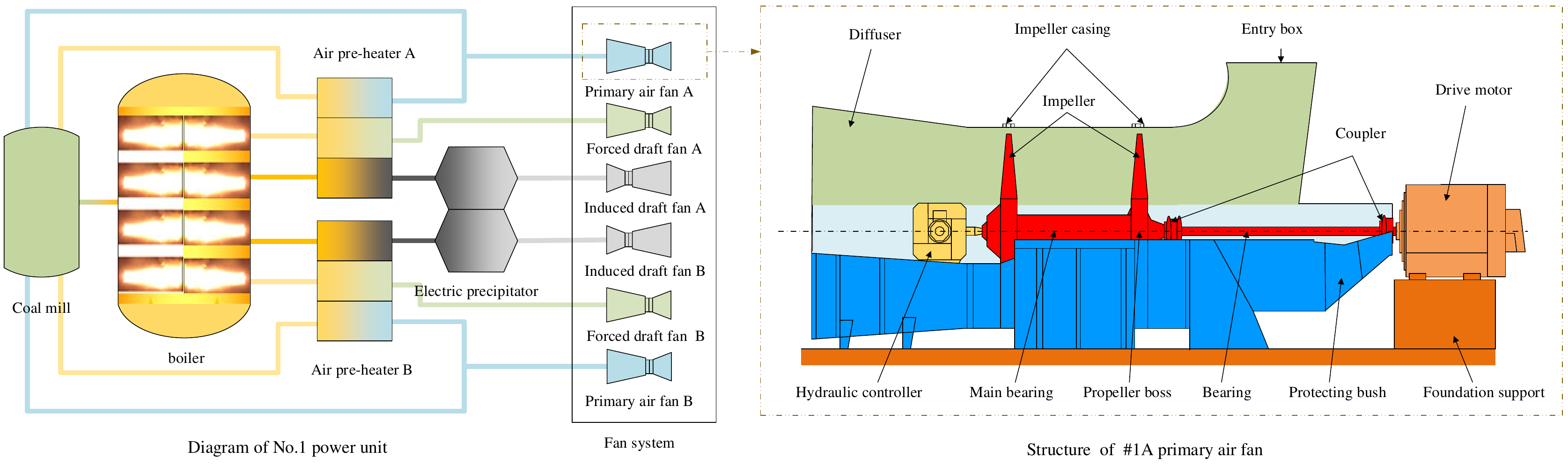}
		\caption{Structure diagram and working condition of the primary air fan.}\label{Structurediagram}
	\end{center}
\end{figure*}
\begin{figure*}[htbp]
	\centering
	
	\subfigure[PCA(CPV=0.85):$T^2$]
	{
		\centering
		\includegraphics[width=2.3in]{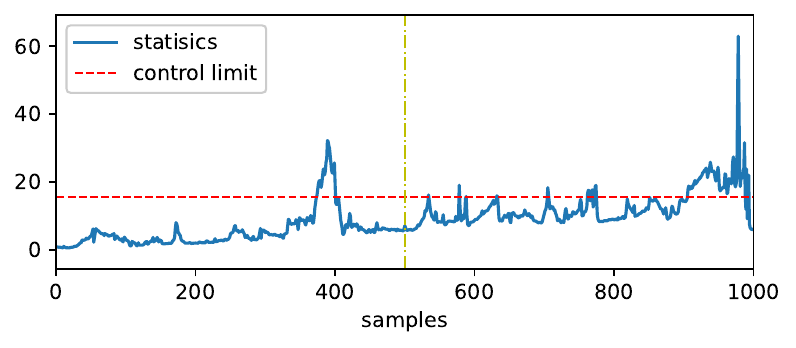}
	}%
	\subfigure[PCA(CPV=0.85):$Q$]
	{
		\centering
		\includegraphics[width=2.3in]{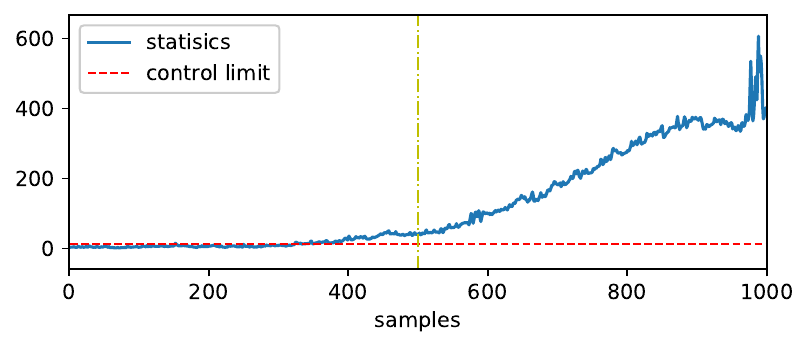}
	}%
	\subfigure[PCA(CPV=0.90):$T^2$]
	{
		\centering
		\includegraphics[width=2.3in]{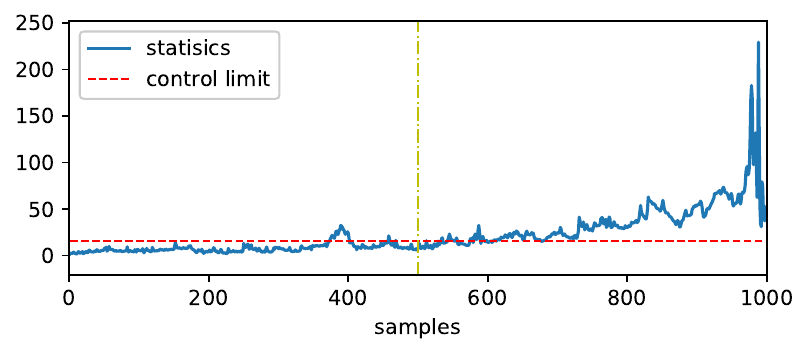}
	}%
	
	\subfigure[PCA(CPV=0.90):$Q$]
	{
		\centering
		\includegraphics[width=2.3in]{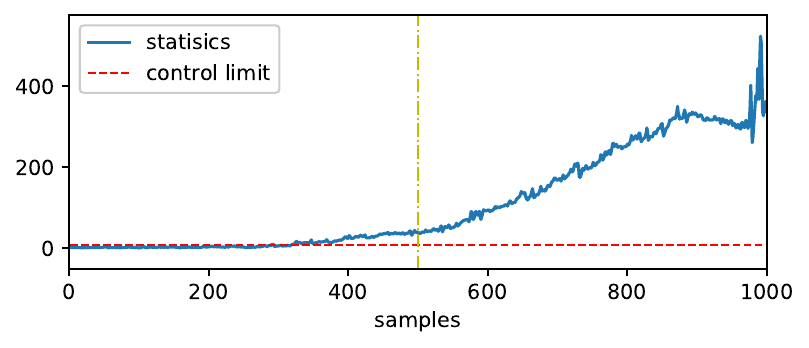}
	}%
	\subfigure[PCA(CPV=0.95):$T^2$]{
		\centering
		\includegraphics[width=2.3in]{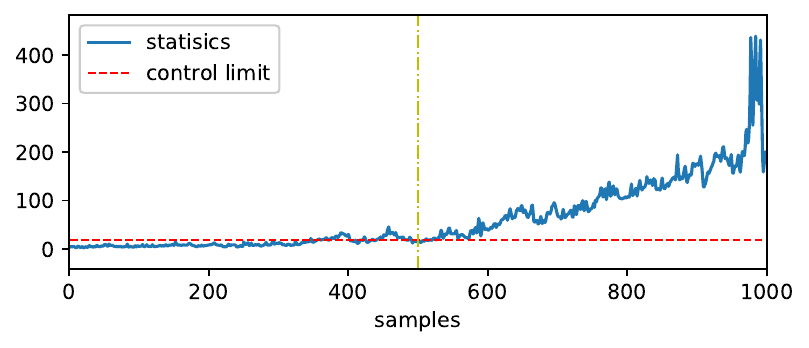}
	}%
	\subfigure[PCA(CPV=0.95):$Q$]
	{
		\centering
		\includegraphics[width=2.3in]{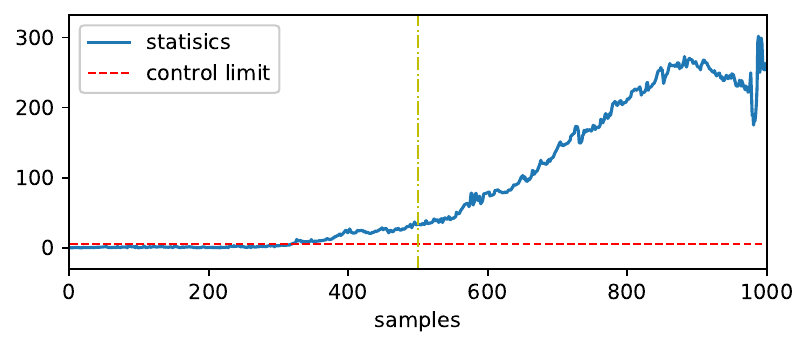}
	}%
	
	\centering
	\caption{Detection performance of PCA in the fan system.}
	\label{monitoringchartsofplantdataPCA}
\end{figure*}

According to the above parameters, 100 independent repeated experiments were conducted. Some classic methods, such as PCA \cite{1991Multivariate,1996Identification}, DPCA \cite{1995Disturbance}, ICA \cite{2004Statistical}, MD \cite{2019Incipient} are used to verifying the effectiveness of HVM. For PCA and DPCA, the cumulative percent variance (CPV) is 0.80. Generally speaking, a larger CPV leads to a larger number of principal components. The number of principal components is 4 in PCA when CPV is 0.80. In order to be consistent, the CPV is also 80\% in DPCA. The number of principal components is 12 (the total dimension is 15). The time lag in DPCA is 2 \cite{KU1995179}. The number of independent components (IC) in ICA equals to 3.  The means of false alarm rates (FARs) and fault detection rates (FDRs) are listed in Table \ref{ThemeansofFARsandFDRs}. The statistics of mentioned methods in experiment \textrm{II} are depicted in Fig. \ref{fault1Simu}. In experiment \textrm{I}, the statistical characteristics of both continuous and binary variables  are slightly different under normal and faulty conditions. The FDR of $Q$ in DPCA is 35.21\%. When the information carried in both continuous and binary variables is simultaneously mined, the FDR is improved to 52.34\%. The difference in the distributions of continuous variables under normal and fault conditions is narrowed, and the difference of binary variables is more significant in experiment \textrm{II}. The best FDR of traditional methods with continuous variables is just 6.47\%. But the FDR of HVM is 94.73\%, and the FAR is only 0.60\%.

\subsection{Fan system of the power plant}
The ultra-supercritical thermal power plants have made great contributions to the development of society and still play a pivotal role in the current power system \cite{QinYihao2019Data}. Efficient process monitoring is the foundation of continuous and stable operation for power plants. In this case, the effectiveness and efficiency of PVM is verified by an actual data collected from Zhoushan Power Plant, Zhejiang Province, China. The number of variables monitoring the No.1 power unit in Zhoushan power plant is more than $17380$, and the number of binary variables among them is as many as $8820$ \cite{wang2020anomaly}. In the fan system of the No.1 power unit, $260$ continuous variables and $495$ binary variables are collected, where the number of binary variables is more than that of continuous variables \cite{MinWang2021JASfinal}.
\begin{figure*}[htbp]
	\centering
	
	\subfigure[DPCA(CPV=0.85):$T^2$]
	{
		\centering
		\includegraphics[width=2.3in]{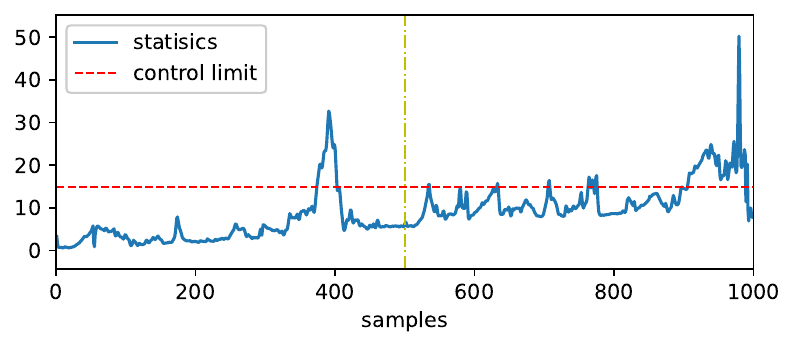}
	}%
	\subfigure[DPCA(CPV=0.85):$Q$]
	{
		\centering
		\includegraphics[width=2.3in]{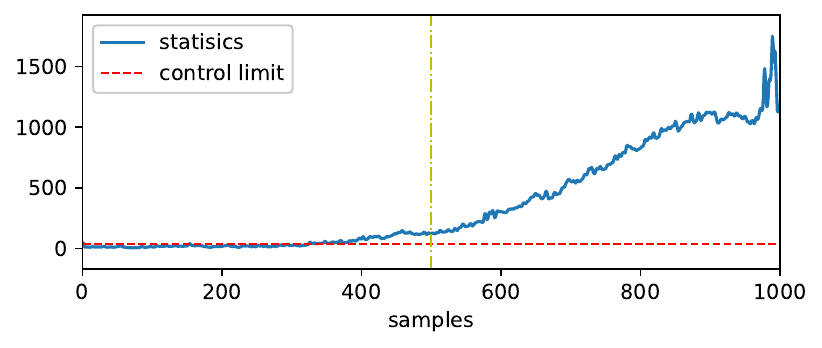}
	}%
	\subfigure[DPCA(CPV=0.90):$T^2$]
	{
		\centering
		\includegraphics[width=2.3in]{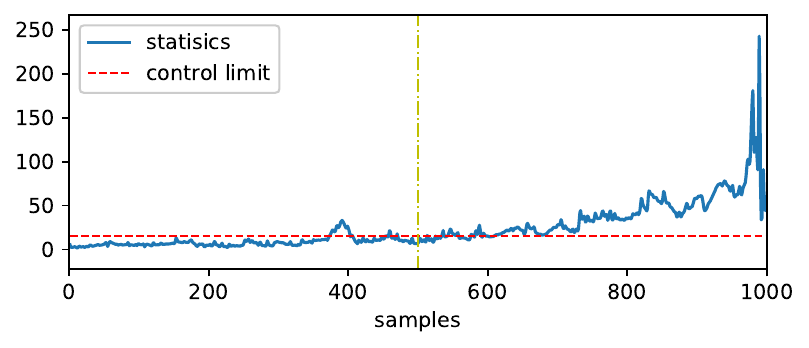}
	}%
	
	\subfigure[DPCA(CPV=0.90):$Q$]
	{
		\centering
		\includegraphics[width=2.3in]{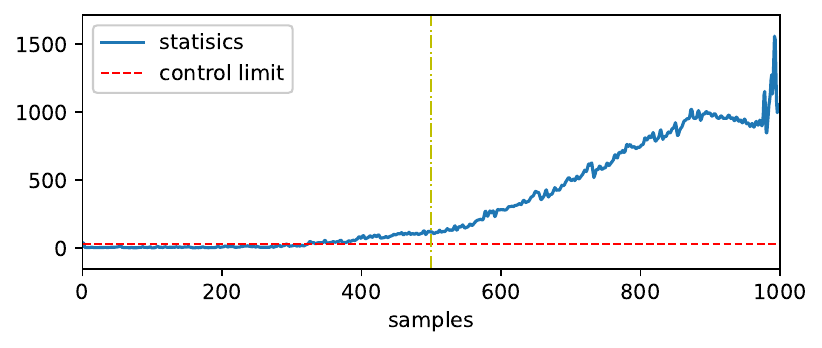}
	}%
	\subfigure[DPCA(CPV=0.95):$T^2$]
	{
		\centering
		\includegraphics[width=2.3in]{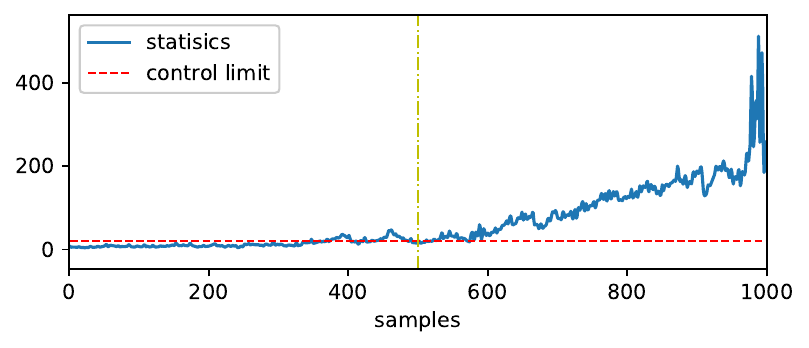}
	}%
	\subfigure[DPCA(CPV=0.95):$Q$]
	{			
		\centering
		\includegraphics[width=2.3in]{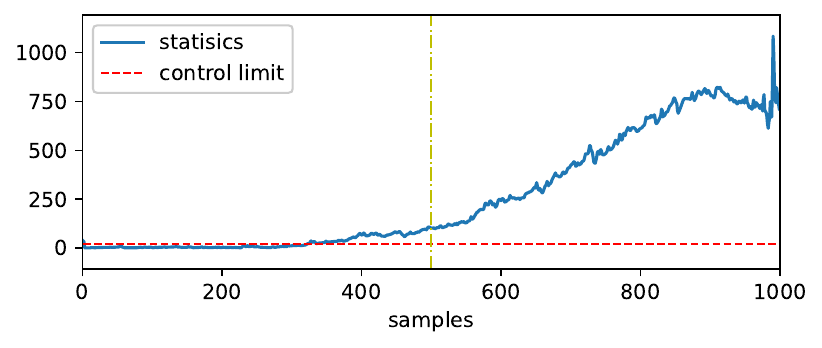}
	}
	
	\centering
	\caption{Detection performance of DPCA in the fan system.}
	\label{monitoringchartsofplantdataDPCA}
\end{figure*}
\begin{table*}[]
	\scriptsize
	\centering
	\caption{The FARs and FDRs in the fan system}
	\label{Thesimulationexperimentresults}
	\scalebox{0.95}[0.95]{
		\begin{tabular}{ccccccccccccccccccc}
			\toprule
			\multirow{3}{*}{methods}
			& \multicolumn{2}{c}{PCA(CPV=0.80)} & \multicolumn{2}{c}{PCA(CPV=0.85)} & \multicolumn{2}{c}{PCA(CPV=0.90)} & \multicolumn{2}{c}{PCA(CPV=0.95)} & \multicolumn{2}{c}{DPCA(CPV=0.80)} & \multicolumn{2}{c}{DPCA(CPV=0.85)} \\
			
			\cmidrule(r){2-3} \cmidrule(r){4-5} \cmidrule(r){6-7} \cmidrule(r){8-9} \cmidrule(r){10-11} \cmidrule(r){12-13}
			
			&$T^2$ &$Q$ &$T^2$ &$Q$ &$T^2$ &$Q$ &$T^2$ &$Q$ &$T^2$ &$Q$ &$T^2$ &$Q$ \\
			
			\midrule
			
			FAR(\%) &3.80 &34.40 &5.40 &35.60 &6.20  &37.20  &22.40  &36.40  &3.80 &33.80  &5.80  &35.20  \\
			
			FDR(\%) &19.40 &100.00 &19.80 &100.00 &83.80  &100.00  &97.60  &100.00  &17.20 &100.00  &20.20  &100.00\\
			
			\midrule
			\midrule
			
			\multirow{3}{*}{methods}
			& \multicolumn{2}{c}{DPCA(CPV=0.90)} & \multicolumn{2}{c}{DPCA(CPV=0.95)} & \multicolumn{3}{c}{ICA(IC=10)} & \multicolumn{3}{c}{ICA(IC=15)} & \multicolumn{1}{c}{\multirow{3}{*}{MD}} & \multicolumn{1}{c}{\multirow{3}{*}{HVM}}\\
			
			\cmidrule(r){2-3} \cmidrule(r){4-5} \cmidrule(r){6-8} \cmidrule(r){9-11}
			
			&$T^2$ &$Q$ &$T^2$ &$Q$ &$I^2$ &$I_e^2$&$Q$ &$I^2$ &$I_e^2$&$Q$\\
			
			\midrule
			
			FAR(\%)  &7.80  &36.00  &19.60  &36.20  &39.80 &42.00  &39.60 &42.10  &38.60 &39.80  &42.00 & 4.20 \\
			
			FDR(\%)  &86.60  &100.00  &95.60  &100.00  &100.00 &100.00  &99.80 &100.00  &100.00 &100.00 &100.00 & 100.00\\
			
			\bottomrule	
	\end{tabular} }
\end{table*}
\begin{table}[tp]
	\scriptsize
	\centering
	\caption{The number of principal components in PCA and DPCA.}
	\label{principalcomponentPCADPCA}
	\begin{tabular}{m{1.5cm}<{\centering}m{0.3cm}<{\centering}m{0.4cm}<{\centering}m{0.4cm}<{\centering}m{0.4cm}<{\centering}m{0.3cm}<{\centering}m{0.4cm}<{\centering}m{0.3cm}<{\centering}m{0.5cm}<{\centering}m{0.8cm}<{\centering}}
		\toprule
		\multirow{1}{*}{methods} & \multicolumn{4}{c}{PCA} & \multicolumn{4}{c}{DPCA}  \\
		
		\cmidrule(r){2-5} \cmidrule(r){6-9}
		
		CPV(\%) &80 &85 &90 &95 &80 &85 &90 &95\\
		
		\midrule
		
		number &2 &3 &5 &7 &2 &3 &5 &8 \\
		
		\bottomrule	
	\end{tabular}
\end{table}

A vibration fault of the \#1A primary air fan in the No.1 power unit occurred on September 3, 2017. The primary air fan is the driving force for the transportation of pulverized coal, and provides hot air for the drying and oxygen for the combustion of pulverized coal. The working environment and structure diagram of the primary air fan are shown in Fig. \ref{Structurediagram}. According to the recommendation of the practical engineers, $35$ continuous variables and $35$ binary variables are sampled every $5$ seconds. $1000$ instances under normal condition are used for modeling. $500$ samples before and after the fault are collected respectively to test the effectiveness and efficiency. 
\begin{figure*}[htbp]
	\centering
	
	\subfigure[MD]
	{
		\centering
		\includegraphics[width=2.3in]{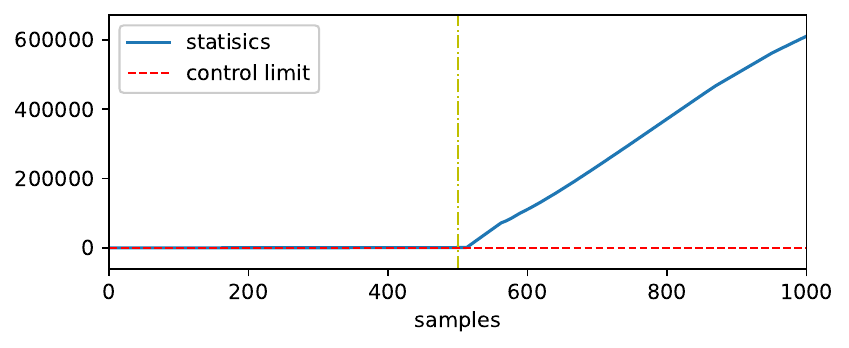}
		\label{monitoringchartsofplantdataMD}
	}%
	\subfigure[ICA(IC=15):$I^2$]
	{
		\centering
		\includegraphics[width=2.3in]{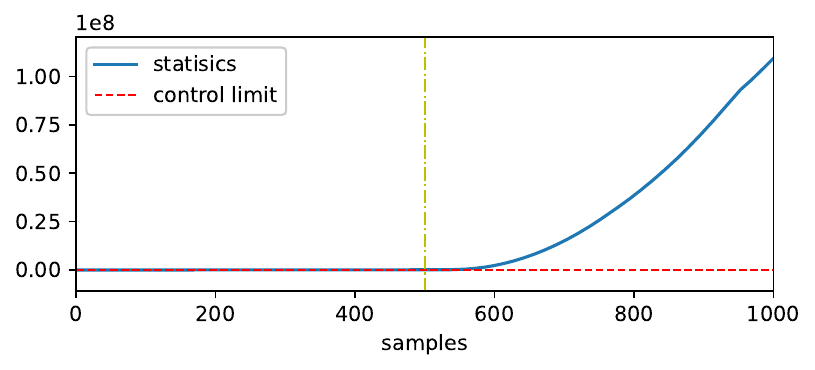}
		\label{monitoringchartsofplantdataICAI2}
	}%
	\subfigure[ICA(IC=15):$I_e^2$]
	{
		\centering
		\includegraphics[width=2.3in]{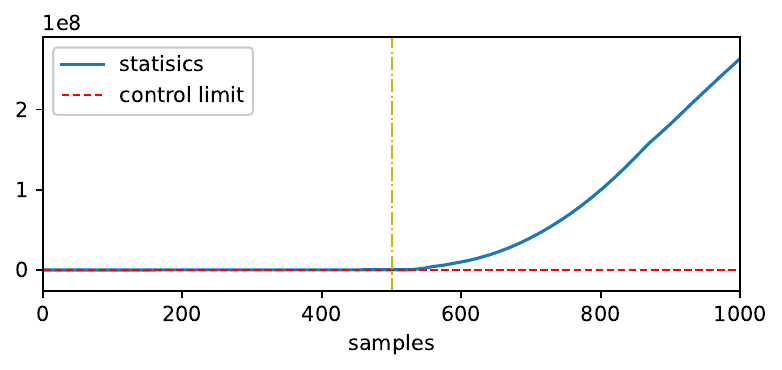}
		\label{monitoringchartsofplantdataICAIe2}
	}%
	
	\subfigure[MD(Logarithm of statistic)]
	{
		\centering
		\includegraphics[width=2.3in]{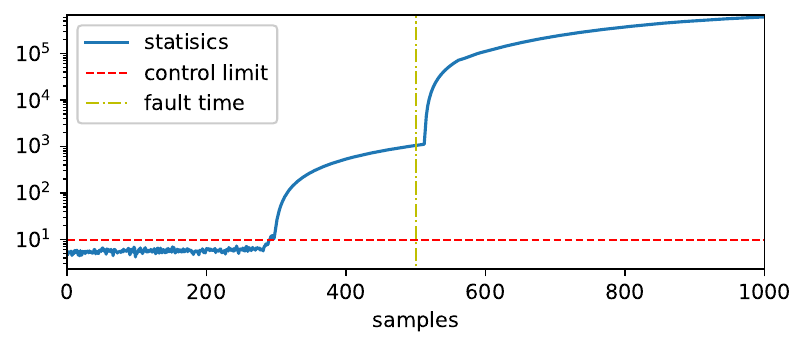}
		\label{monitoringchartsofplantdataMDlog}
	}%
	\subfigure[ICA(IC=15):$I^2$(Logarithm of statistic)]
	{
		\centering
		\includegraphics[width=2.3in]{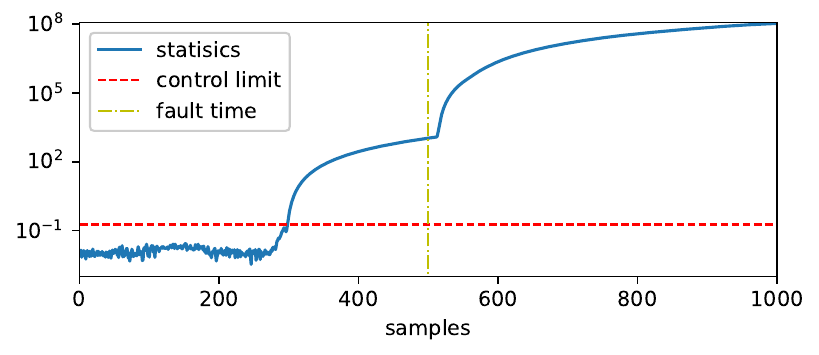}
		\label{monitoringchartsofplantdataICAI2log}
	}%
	\subfigure[ICA(IC=15):$I_e^2$(Logarithm of statistic)]
	{
		\centering
		\includegraphics[width=2.3in]{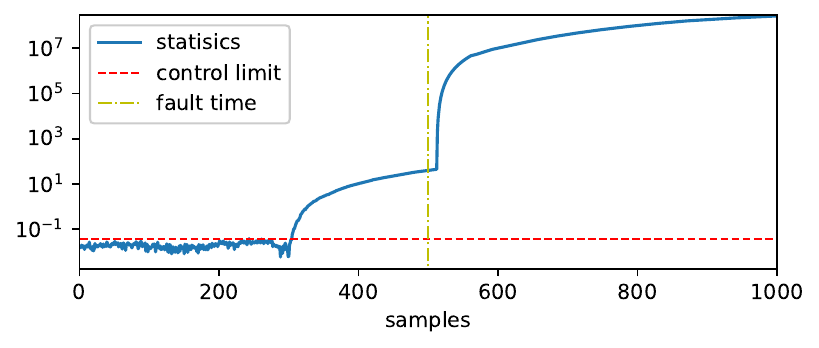}
		\label{monitoringchartsofplantdataICAIe2log}
	}%
	
	\subfigure[ICA(IC=15):$Q$]
	{
		\centering
		\includegraphics[width=2.3in]{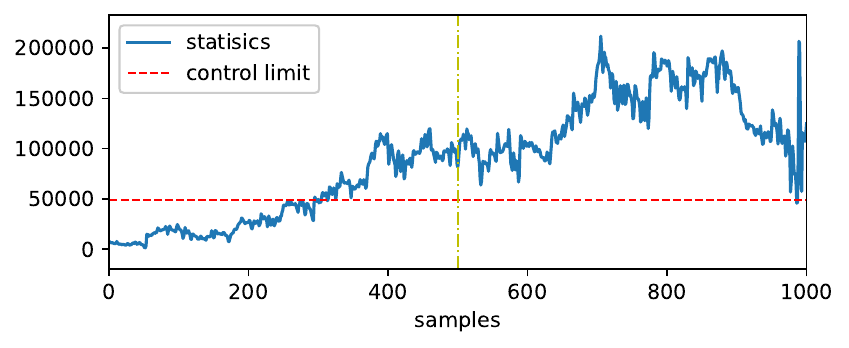}
		\label{monitoringchartsofplantdataICAQ}
	}%
	\subfigure[HVM]
	{
		\centering
		\includegraphics[width=2.3in]{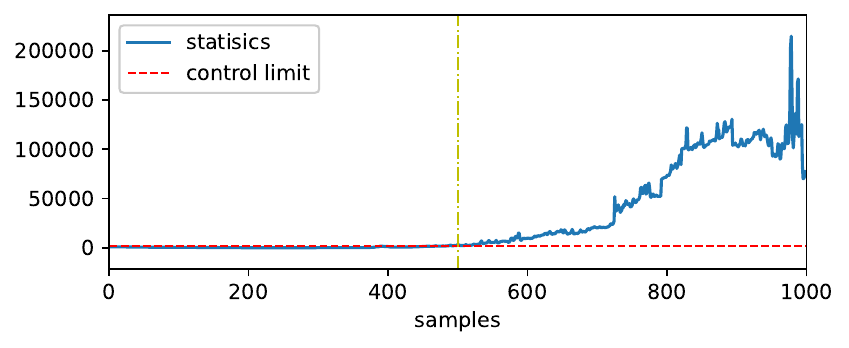}
		\label{monitoringchartsofplantdataHVM}
	}%
	\subfigure[HVM(Logarithm of statistic)]
	{
		\centering
		\includegraphics[width=2.3in]{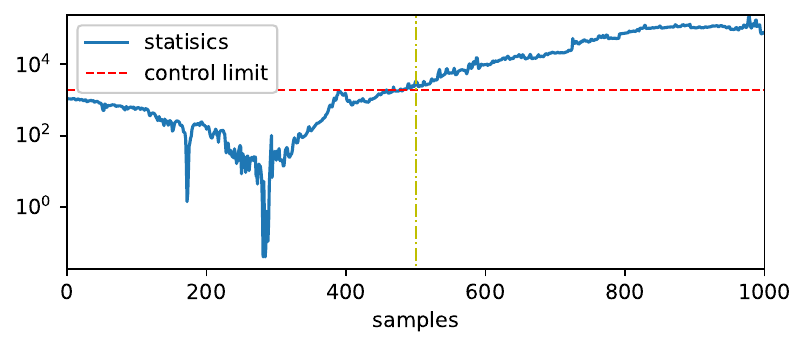}
		\label{monitoringchartsofplantdataHVMlog}
	}%
	
	\centering
	\caption{Detection performance of MD, ICA and HVM in the fan system.}
	\label{monitoringchartsofplantdata}
\end{figure*}

For traditional monitoring models, PCA \cite{1991Multivariate,1996Identification}, DPCA \cite{1995Disturbance}, ICA \cite{2004Statistical} and MD \cite{2019Incipient} are adopted for process monitoring with continuous variables. Experiments were conducted with CPV equals to $0.80$, $0.85$, $0.90$, $0.95$ respectively for PCA and DPCA, where $T^2$ and $Q$ statistic are calculated. The number of principal components in PCA and DPCA is listed in Table \ref{principalcomponentPCADPCA}. For DPCA, the time lag is $2$ \cite{KU1995179}. The FARs and FDRs of $T^2$ statistics keep increasing with the increase of CPV for PCA and DPCA. The best results appear on $T^2$ statistics of PCA and DPCA at CPV=0.9. The FAR and FDR of PCA with CPV=0.9 are 6.20\% and 83.80\% respectively. For DPCA, the FAR and FDR with CPV=0.9 are 7.80\% and 86.70\% respectively. The monitoring charts of PCA with CPV equals to  $0.85$, $0.90$ and $0.95$ are shown in Fig. \ref{monitoringchartsofplantdataPCA}. The statistics of DPCA when CPV is $0.85$, $0.90$ and $0.95$ are depicted in Fig. \ref{monitoringchartsofplantdataDPCA}. In ICA, IC=10 and IC=15 are considered. The results show that the monitoring performances of ICA are similar with different IC. The statistics of ICA when IC=15 is shown in Fig. \ref{monitoringchartsofplantdataICAI2}, \ref{monitoringchartsofplantdataICAIe2} and \ref{monitoringchartsofplantdataICAQ}. The detection performance of MD can be seen in Fig. \ref{monitoringchartsofplantdataMD}. The logarithmic statistics of $I^2, I_e^2$ in ICA and MD are shown in Fig. \ref{monitoringchartsofplantdataMDlog}, \ref{monitoringchartsofplantdataICAI2log} and \ref{monitoringchartsofplantdataICAIe2log}. The FDRs of MD and ICA are satisfactory, but the FARs is too high to be accepted. However, the FDR of HVM is 100\% and FAR of HVM is 4.2\% when both continuous and binary variables are utilized. Continuous variables contain current, air volume, vibration, temperature \textit{etc.} of the fans. Binary variables mainly including control command signal, vibration over-limit signal, bearing vibration danger signal, moving blade position feedback signal, state signal, \textit{etc.} are taken into consideration in HVM. Variables that are more strongly correlated with other variables tend to change easily when any other variable changes. In this case, the vibration-related variables have relatively larger weights. The detection performance of HVM is depicted in Fig. \ref{monitoringchartsofplantdataHVM} and \ref{monitoringchartsofplantdataHVMlog}. The FARs and FDRs of all methods are listed in Table \ref{Thesimulationexperimentresults}.  

\section{Conclusions}
\label{Conclusion}
This paper focuses on the issue of hybrid variable monitoring only based on healthy state data and proposes a novel unsupervised process monitoring framework for hybrid variables named PVM. The statistics suitable for hybrid variables are defined and the physical explanation behind the framework is elaborated. In addition, the estimation of parameters is derived in detail and the detectable conditions of HVM is analyzed. Finally a numerical simulation and an actual case in the plant process of thermal power are utilized to verified the effectiveness and efficiency of the proposed model. Studies demonstrate that HVM have the superiority when the information of both continuous and binary variables are  effectively utilized.

\begin{ack}   
This work was supported by the National Natural Science Foundation of China under Grant  62033008, 61873143.
\end{ack}

\appendix
\section{Proof of Theorem \ref{theoremrelationship}}
\label{appendix1}

For continuous variables $x^j$ and $ x^{j'}$, the joint probability function of $x^j$ and $ x^{j'}$ is
\begin{align}\label{Thejointprobabilityfunctionxjxjj}
	&f(x^j, x^{j'}) = (2\pi \sigma^j \sigma^{j'})^{-1}(1-\rho^{2})^{-1/2} \exp \{-(2-2\rho^{2})^{-1} \notag \\
	&~~~~~~~~~~~\times[(x^j-\mu^j)^{2}(\sigma^j)^{-2}+(x^{j'}-\mu^{j'})^{2}(\sigma^{j'})^{-2}\notag \\
	&~~~~~~~~~~~-2\rho(x^j-\mu^j)(x^{j'}-\mu^{j'}) (\sigma^{j})^{-1} (\sigma^{j'})^{-1}  ] \} .
\end{align}
Then $P( {x'}^j=1, {x'}^{j'} = 1 )$ is learned as
\begin{align}\label{Exjxjj}
	&P( {x'}^j=1, {x'}^{j'} = 1 )= P( x^j > \mu^{j},x^{j'} > \mu^{j'})\notag \\
	& =\int\limits_{\mu^{j}}^\infty  {\int\limits_{\mu^{j'}}^\infty  f( x^j,x^{j'}) }d{x^j}d{x^{j'}} =\int\limits_{0}^\infty  {\int\limits_{0}^\infty  f( {y^j},{y^{j'}}) }d{y^j}d{y^{j'}} ,
\end{align}
where $y^j=(x^j-\mu^j)(\sigma^j)^{-1}$ and $y^{j'}=(x^{j'}-\mu^{j'}){\sigma^{j'}}^{-1}$. Since 
\begin{align}\label{Thejointprobabilityfunctionxjxjj1}
	f(y^j,y^{j'}) &= (2\pi)^{-1}(1-\rho^{2})^{-1/2} \exp \{-(2-2\rho^{2})^{-1} \notag \\
	&\times[(y^j)^{2} +(y^{j'})^{2}-2\rho y^j y^{j'} ] \} .
\end{align}
Thus
\begin{align}\label{Exjxjj11}
	&P( {x'}^j=1, {x'}^{j'} = 1 ) =\int\limits_{0}^\infty  \int\limits_{0}^\infty \{(2\pi)^{-1}(1-\rho^{2})^{-1/2} \notag \\
	&\times\exp \{-(2-2\rho^{2})^{-1} [(y^j)^{2} +(y^{j'})^{2}-2\rho y^j y^{j'} ] \} \} d{y^j}d{y^{j'}}\notag \\
	&=\int\limits_{0}^\infty  \int\limits_{0}^{\pi/2} \{(2\pi)^{-1}(1-\rho^{2})^{-1/2}r \notag \\
	&\times \exp \{-(2-2\rho^{2})^{-1}(1-\rho\sin 2\alpha) \}\}d\alpha dr\notag \\
	&=\int\limits_{0}^{\pi/2} \{(2\pi)^{-1}(1-\rho^{2})^{-1/2} (1-\rho\sin 2\alpha)^{-1}\}d\alpha \notag \\
	&=\int\limits_{0}^{\pi/2} \{(2\pi)^{-1}(1-\rho^{2})^{-1/2} \notag \\
	&\times(1+\tan \alpha^2-2\rho\tan \alpha)^{-1}\}d\tan \alpha \notag \\
	&= \frac{1}{{2\pi }}\arcsin \rho+ 0.25.
\end{align}

In the same way, we have
\begin{align}\label{Exjxjj11}
	&P( {x'}^j=0, {x'}^{j'} = 0 ) =P( x^j \leq \mu^{j},x^{j'} \leq \mu^{j'})\notag \\
	& =\int\limits_{-\infty}^{\mu^{j}}  {\int\limits_{-\infty}^{\mu^{j}}  f( x^j,x^{j'}) }d{x^j}d{x^{j'}} =\int\limits_{-\infty}^0  {\int\limits_{-\infty}^0  f( {y^j},{y^{j'}}) }d{y^j}d{y^{j'}} \notag \\
	&=\int\limits_{-\infty}^0  \int\limits_{-\infty}^0 \{(2\pi)^{-1}(1-\rho^{2})^{-1/2} \exp \{-(2-2\rho^{2})^{-1}\notag \\
	&\times [(y^j)^{2} +(y^{j'})^{2}-2\rho y^j y^{j'} ] \} \} d{y^j}d{y^{j'}}.
\end{align}

Since $y^j$ and $y^{j'}$ are Gaussian distributions, it can be obtained that 
\begin{align}\label{Thejointprobabilityfunctionzjzj1}
	f(z^j,z^{j'})=f(-y^j,-y^{j'}) = f(y^j,y^{j'}).
\end{align}
where $z^j=-y^j$, $z^{j'}=-y^{j'}$. Then
\begin{align}\label{Exjxjj111}
	&P( {x'}^j=0, {x'}^{j'} = 0 )=\int\limits_{0}^\infty  \int\limits_{0}^\infty \{(2\pi)^{-1}(1-\rho^{2})^{-1/2} \notag \\
	&\times \exp \{-(2-2\rho^{2})^{-1} [(z^j)^{2} +(z^{j'})^{2}-2\rho z^j z^{j'} ] \} \} d{z^j}d{z^{j'}}\notag \\
	&= \frac{1}{{2\pi }}\arcsin \rho+ 0.25.
\end{align}

The MI $\mathcal M({x'}^j, {x'}^{j'})$ of ${x'}^j$ and ${x'}^{j'}$ ( ${x'}^j$ and ${x'}^{j'}$ are constructed through equation \eqref{clippingprocessing}) is defined as
\begin{align}\label{computingofMI}
	\mathcal M({x'}^j, {x'}^{j'}) = \sum\limits_{{{x'}^j},{{x'}^{j'}}} {P( {{{x'}^j},{{x'}^{j'}}})\log \frac{{P( {{{x'}^j},{{x'}^{j'}}})}}{{P( {{{x'}^j}})P( {{{x'}^{j'}}})}}}.
\end{align}

Since $x^j$ is a Gaussian process, it is obvious that $P( {x'}^j=1)= \int\limits_{\mu^{j}}^\infty x^j d{x^j}=1/2$. In the same way, we have $P( {{{x'}^j}}) = P( {{{x'}^{j'}}})=1/2$. Then equation \eqref{computingofMI} is 
\begin{align}\label{computingofMI4}
	&\mathcal M({x'}^j, {x'}^{j'}) = \sum\limits_{{x'}^j,{x'}^{j'}} P( {x'}^j,{x'}^{j'})\log 4 P( {{{x'}^j},{{x'}^{j'}}})\notag \\
	&=P( {x'}^j=0,{x'}^{j'}=0)\log 4 P( {x'}^j=0,{x'}^{j'}=0)\notag \\
	&+ P( {x'}^j=0,{x'}^{j'}=1)\log 4 P( {x'}^j=0,{x'}^{j'}=1)\notag \\
	&+ P( {x'}^j=1,{x'}^{j'}=0)\log 4 P( {x'}^j=1,{x'}^{j'}=0)\notag \\
	&+ P( {x'}^j=1,{x'}^{j'}=1)\log 4 P( {x'}^j=1,{x'}^{j'}=1),
\end{align}

Let $P( x^j = 0| x^{j'}=0) = \lambda, P( x^j = 1| x^{j'}=1) = \lambda'$. Since $P(x^{j'}=0)=P(x^{j'}=1)=1/2$, then
\begin{align}\label{computingofMIlambda}
	&P( {x'}^j=0,{x'}^{j'}=0)=\frac{1}{2} \lambda, \\
	&P( {x'}^j=0,{x'}^{j'}=1)=\frac{1}{2} (1-\lambda'), \\
	&P( {x'}^j=1,{x'}^{j'}=0)=\frac{1}{2} (1-\lambda), \\
	&P( {x'}^j=1,{x'}^{j'}=1)=\frac{1}{2} \lambda '.
\end{align}

According to equation \eqref{Exjxjj11} and \eqref{Exjxjj111}, we have 
\begin{align}\label{lambdalambda1}
	\lambda = \lambda'= \frac{1}{{\pi }}\arcsin \rho+ 0.5.
\end{align}

Then 
\begin{align}\label{lambdaMI}
	&\mathcal M({x'}^j, {x'}^{j'}) =2 P( {x'}^j=0,{x'}^{j'}=0)\log 4 P( {x'}^j=0,{x'}^{j'}=0)\notag \\
	&+ 2P( {x'}^j=0,{x'}^{j'}=1)\log 4 P( {x'}^j=0,{x'}^{j'}=1)\notag \\
	&=\lambda \log 2\lambda+ (1-\lambda)\log 2(1-\lambda)\notag \\
	&=(\frac{1}{{\pi }}\arcsin \rho+ 0.5) \log (\frac{2}{{\pi }}\arcsin \rho+ 1)\notag \\
	&+ (0.5-\frac{1}{{\pi }}\arcsin \rho)\log (1-\frac{2}{{\pi }}\arcsin \rho),
\end{align}

According to equation \eqref{lambdaMI} and lemma \ref{lemma1}, it is learned that
\begin{align}\label{lambdaMI1}
	&\mathcal M({x'}^j, {x'}^{j'}) =(\frac{1}{{\pi }}\arcsin \rho+ 0.5) \log (\frac{2}{{\pi }}\arcsin \rho+ 1)\notag \\
	&+ (0.5-\frac{1}{{\pi }}\arcsin \rho)\log (1-\frac{2}{{\pi }}\arcsin \rho),
\end{align}
where $\rho =  [1-e^{-2\mathcal M(x^j, x^{j'})}]^{1/2}$.\qed

\section{Proof of Proposition \ref{propositionPxjxj}} 
\label{appendixProofproposition3} 
	Let
\begin{align}\label{pxj1pxj11}
	P( x^j = 1| x^{j'}=1) = \varsigma, P( x^j = 1| x^{j'}=0) = \varsigma' ,
\end{align}
it has
\begin{align}\label{pxj0pxj11}
	&P( x^j = 0| x^{j'}=1) = 1-\varsigma ,\\
	\label{pxj0pxj10}
	&P( x^j = 0| x^{j'}=0) = 1-\varsigma' .
\end{align}
Then we have 
\begin{align}\label{thelikelihoodfunctionofvarphin1}
	&P( x^j = \psi_{x^j}| x^{j'} = \psi_{x^{j'}})=\varsigma^{\psi_{x^j}\psi_{x^{j'}}} (1-\varsigma)^{\psi_{x^{j'}}-\psi_{x^j}\psi_{x^{j'}}} \notag \\
	&~~~~~~~~\times \varsigma'^{\psi_{x^j}-\psi_{x^j}\psi_{x^{j'}}}(1-\varsigma')^{1+\psi_{x^j}\psi_{x^{j'}}-\psi_{x^j}-\psi_{x^{j'}}}.
\end{align}

Since
\begin{align}\label{Pxjxj1denoted1}
	&P(x^j,x^{j'})=P(x^{j}=\psi_{x^j},x^{j'}=\psi_{x^{j'}}) \notag \\
	&=P(x^{j'}=\psi_{x^{j'}})P(x^{j}=\psi_{x^j}|x^{j'}=\psi_{x^{j'}}).
\end{align}
Thus Proposition \ref{propositionPxjxj} is proved. \qed

\section{Proof of Theorem \ref{theorempxjxj1}} 
\label{appendixProofTheoremtheorempxjxj1} 
Since 
\begin{align}\label{hatvarphinjoindis1}
	&P( x^j = x_1^j, x^{j'} = x_1^{j'}) \ldots P( x^j = x_n^j, x^{j'} = x_n^{j'}) \notag \\
	& = \prod\limits_{i = 1}^{n} P( x^{j'} = x_i^{j'}){P( x^j = x_i^j| x^{j'} = x_i^{j'} )}.
\end{align}

The likelihood function is 
\begin{align}\label{thelikelihoodfunctionofvarphi1}
	\ell( {\varsigma ,\varsigma'} ) &= \prod\limits_{i = 1}^{n} P( x^{j'} = x_i^{j'}){P( x^j = x_i^j| x^{j'} = x_i^{j'} )}  \notag \\
	&= \prod\limits_{i = 1}^{n} P( x^{j'} = x_i^{j'}) \prod\limits_{i = 1}^{n} {P( x^j = x_i^j| x^{j'} = x_i^{j'} )}  \notag \\
	&= \varpi {\varsigma ^{\sum\limits_{i = 1}^{n} {x_i^j x_i^{j'}} }}{( {1 - \varsigma } )^{\sum\limits_{i = 1}^{n} {x_i^{j'}}  - \sum\limits_{i = 1}^{n} x_i^j x_i^{j'} }} {{\varsigma '}^{\sum\limits_{i = 1}^{n} x_i^j - \sum\limits_{i = 1}^{n} x_i^j x_i^{j'}  }}  \notag \\
	&\times {\left( {1 - \varsigma '} \right)^{n + \sum\limits_{i = 1}^{n} x_i^j x_i^{j'}  - \sum\limits_{i = 1}^{n} (x_i^j + x_i^{j'}) }},
\end{align}
where $\varpi$ is a constant. 
Then $\frac{\partial \ell ( {\eta ,\eta '} )}{\partial \eta }$ can be obtained as
\begin{align}\label{Thederivative}
	&\frac{\partial \ell ( {\varsigma ,\varsigma '} )}{\partial \varsigma } = \varpi {{\varsigma '}^{\sum\limits_{i = 1}^{n} x_i^j - \sum\limits_{i = 1}^{n} x_i^j x_i^{j'}  }} {\left( {1 - \varsigma'} \right)^{n + \sum\limits_{i = 1}^{n} x_i^j x_i^{j'}  - \sum\limits_{i = 1}^{n} (x_i^j + x_i^{j'}) }} \notag \\
	&[(-1)( {1 - \varsigma } )^{-1}({\sum\limits_{i = 1}^{n} x_i^{j'} - \sum\limits_{i = 1}^{n} x_i^j x_i^{j'} })  {( {1 - \varsigma } )^{\sum\limits_{i = 1}^{n} x_i^{j'}  - \sum\limits_{i = 1}^{n} x_i^j x_i^{j'} }} \notag \\
	&{\varsigma ^{\sum\limits_{i = 1}^{n} x_i^j x_i^{j'} }} + ({\sum\limits_{i = 1}^{n} x_i^j x_i^{j'} }) {\varsigma^{\sum\limits_{i = 1}^{n} {x_i^j x_i^{j'}} }} \varsigma^{-1}  {( {1 - \varsigma } )^{\sum\limits_{i = 1}^{n} {x_i^{j'}}  - \sum\limits_{i = 1}^{n} x_i^j x_i^{j'} }} ].
\end{align}

Let $\frac{\partial \ell ( {\varsigma ,\varsigma '} )}{\partial \varsigma }=0$, it can be obtained that
\begin{align}\label{Thederivativeofvarphi1}
	&\varsigma = (\sum\limits_{i = 1}^{n} {x_i^j x_i^{j'}} )(\sum\limits_{i = 1}^{n} {x_i^{j'}} )^{-1}.
\end{align}

In the same way, $\frac{\partial \ell ( {\varsigma ,\varsigma '} )}{\partial \varsigma' }$ is
\begin{align}\label{Thederivative1}
	&\frac{\partial \ell ( {\varsigma ,\varsigma '} )}{\partial \varsigma' } = \varpi {\varsigma ^{\sum\limits_{i = 1}^{n} {x_i^j x_i^{j'}} }}{( {1 - \varsigma } )^{\sum\limits_{i = 1}^{n} {x_i^{j'}}  - \sum\limits_{i = 1}^{n} x_i^j x_i^{j'} }} \notag \\
	&[ (\sum\limits_{i = 1}^{n} x_i^j - \sum\limits_{i = 1}^{n} x_i^j x_i^{j'} )  {\varsigma '}^{-1} {\left( {1 - \varsigma '} \right)^{{n} + \sum\limits_{i = 1}^{n} {x_i^j x_i^{j'}}  - \sum\limits_{i = 1}^{n} {( x_i^j+x_i^{j'} )} }}  \notag \\
	&\times {{\varsigma '}^{\sum\limits_{i = 1}^{n} {x_i^j}  - \sum\limits_{i = 1}^{n} x_i^j x_i^{j'} }} + (-1) ({n} + \sum\limits_{i = 1}^{n} {x_i^j x_i^{j'}}  - \sum\limits_{i = 1}^{n} {( x_i^j+x_i^{j'} )} ) \notag \\
	&\times (1 - \varsigma ')^{-1} {\left( {1 - \varsigma '} \right)^{{n} + \sum\limits_{i = 1}^{n} {x_i^j x_i^{j'}}  - \sum\limits_{i = 1}^{n} {( x_i^j+x_i^{j'} )} }} {{\varsigma '}^{\sum\limits_{i = 1}^{n} {x_i^j} - \sum\limits_{i = 1}^{n} x_i^j x_i^{j'} }} ].
\end{align}

Let $\frac{\partial \ell ( {\varsigma ,\varsigma '} )}{\partial \varsigma' }=0$, $\varsigma'$ can be achieved as
\begin{align}\label{Thederivativeofvarphi11}
	&\varsigma' = (\sum\limits_{i = 1}^{n} {x_i^j}  - \sum\limits_{i = 1}^{n} x_i^jx_i^{j'} )({n}-\sum\limits_{i = 1}^{n} {x_i^{j'} })^{-1}.
\end{align}

According to equation (\ref{pxj1pxj11}), (\ref{pxj0pxj11}) and (\ref{pxj0pxj10}), we have
\begin{align}\label{Thederivativeofvarphi12}
	&P( x^j = \psi_{x^j}| x^{j'} =\psi_{x^{j'}})= \{ 1 - \psi_{x^j} + (2\psi_{x^j} - 1 ) \notag \\
	&\times [ \psi_{x^{j'}}{\varsigma} + ( {1 - \psi_{x^{j'}} } ){\varsigma '} ]\}.
\end{align}

Hence, Theorem \ref{theorempxjxj1} is proven.\qed

\bibliographystyle{plain}
\bibliography{./HVM_ref}

%

\end{document}